\documentclass[11pt]{article}

\usepackage[final]{acl}

\usepackage{times}
\usepackage{latexsym}

\usepackage[T1]{fontenc}

\usepackage[utf8]{inputenc}

\usepackage{microtype}

\usepackage{inconsolata}

\usepackage{graphicx}
\usepackage{paralist}

\usepackage{booktabs} 
\usepackage{tabularx} 
\usepackage{subcaption}   

\usepackage{tabularx}   
\usepackage{booktabs}   
\usepackage{array}      
\usepackage[svgnames,table]{xcolor} 
\usepackage{colortbl}   
\usepackage{threeparttable}
\usepackage{amsmath} 
\usepackage{enumitem}
\usepackage{listings}
\usepackage[most]{tcolorbox}
\usepackage[dvipsnames]{xcolor}
\usepackage{amssymb}
\usepackage{pifont}
\usepackage{graphicx} 
\usepackage{adjustbox}

\definecolor{lightgray}{gray}{0.95}
\definecolor{deepblue}{RGB}{70,130,180}
\definecolor{deepgray}{RGB}{119,136,153}
\lstdefinestyle{prompt}{
    basicstyle=\ttfamily\fontsize{7pt}{8pt}\selectfont,
    frame=none,
    breaklines=true,         
    breakatwhitespace=true,  
    backgroundcolor=\color{lightgray},
    breakatwhitespace=true,
    breakindent=0pt,
    escapeinside={(*@}{@*)},
    numbers=none,
    numbersep=5pt,
    xleftmargin=5pt,
    aboveskip=2pt,
    belowskip=2pt,
}
\tcbset{
  aibox/.style={
    top=10pt,
    colback=white,
    enhanced,
    center,
  }
}
\newtcolorbox{AIbox}[2][]{aibox, title=#2,#1}

\title{Beyond Accuracy: Unveiling Inefficiency Patterns in Tool-Integrated Reasoning}

\author{Qisheng Su$^{1,2}$, Shiting Huang$^{1}$, Zhen Fang$^{1}$, Ziyan Chen$^{1}$, Zehui Chen$^{1}$, Feng Zhao$^{1}$\thanks{Corresponding Author} \\
  $^{1}$MoE Key Lab of BIPC, University of Science and Technology of China \\
  $^{2}$Shanghai Innovation Institute \\
  \texttt{nicksu@mail.ustc.edu.cn} \quad \texttt{fzhao956@ustc.edu.cn} \\}

\begin{document}
\maketitle
\pagestyle{plain}

\begin{abstract}

In real-world Tool-Integrated Reasoning (TIR) scenarios, where LLMs interleave reasoning with external tool calls, a major source of inefficiency is that the toolcalls create pauses between LLM requests and cause KV-Cache eviction, forcing recomputation. Also, the long, unfiltered response returned by external tools inflates the KV-Cache, so each decode step spends more time loading the growing cache and thus becomes steadily slower as context length increases. However, existing efficiency metrics like token counts and toolcall counts fail to capture the real model inference latency. To address this, we introduce PTE (Prefill Token Equivalents), a hardware-aware TIR-efficiency metric that unifies internal reasoning and external tool-use costs while explicitly accounting for non-reusable KV-Cache and long-tool-response scenarios. Validation in a high-concurrency industrial setting indicates that PTE aligns significantly better with wall-clock latency than standard token counts, while maintaining consistent efficiency rankings across diverse hardware profiles. We conduct extensive experiments across five TIR benchmarks, quantify their PTE costs, and identify four inefficiency patterns that appear in TIR. We also discover that trajectories with higher PTE costs tend to have lower reasoning correctness, indicating that simply using more tools does not improve the quality of the answer. The code is available at \url{https://github.com/sqs-ustc/tool-reasoning-framework-PTE}.
\end{abstract}
\section{Introduction}

Large Language Models (LLMs) demonstrate remarkable capabilities in complex tasks via Tool-Integrated Reasoning (TIR)\cite{websailorv2,websailor,tongyideepresearch,tora,mathcoder,torl,battleagent,limiagency}. However, TIR efficiency evaluation remains underexplored. In real-world scenarios, toolcalls create pauses between LLM requests that cause KV-Cache eviction\cite{Continuum,KVFlow}, while long, unfiltered tool responses inflate the context length. This renders the memory-bound decode phase significantly more expensive, yet existing TIR benchmarks focus primarily on accuracy\cite{browsecomp,widesearch}, and efficiency metrics rely on naive token counts or toolcall counts\cite{mcpradar,otc} that fail to reflect the true runtime drain.

\begin{figure}[t]
    \centering
    \includegraphics[width=1\linewidth]{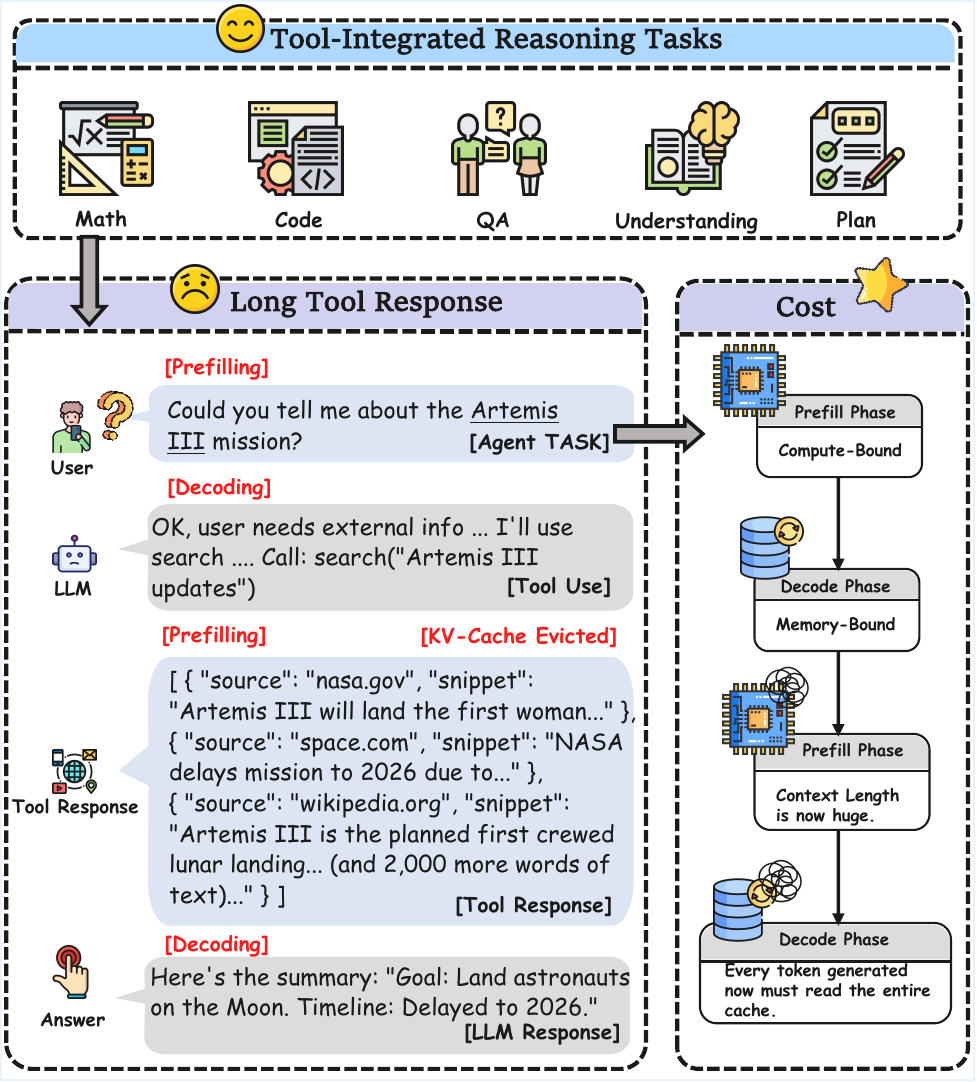}    
        \vspace{-1.5em}
\caption{
Illustration of the asymmetric costs in Tool-Integrated Reasoning (TIR). Toolcalls induce KV-Cache eviction, and long tool responses inflate the context length, significantly increasing the cost of subsequent decoding steps.
} 
    \label{fig:PTEmain}
    \vspace{-0.5em}
\end{figure}

\begin{figure*}[t]
    \centering
    \includegraphics[width=1\textwidth]{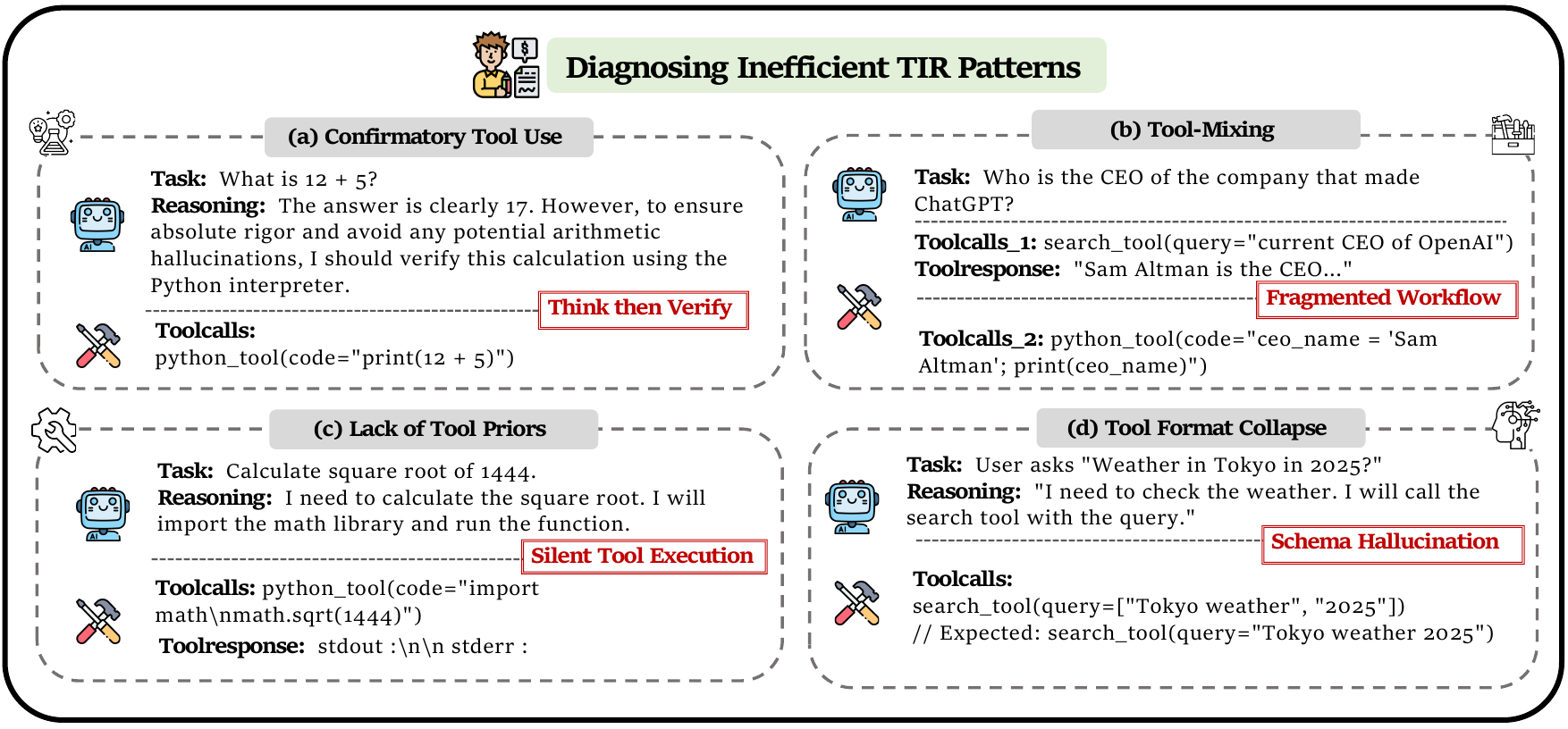}    
        \vspace{-1.5em}
\caption{Overview of the four inefficiency patterns in Tool-Integrated Reasoning.} 
    \label{fig:inefficient_patterns}
    \vspace{-0.5em}
\end{figure*}

Specifically, current efficiency metrics fail to capture the asymmetric costs of the compute-bound prefill phase and the memory-bound decode phase. This oversight proves costly in TIR, where toolcalls trigger KV-Cache eviction and long responses inflate the context, directly increasing the HBM transfer overhead during decode for every subsequent reasoning token. Consequently, existing metrics lack a unified framework to weigh the true expense of internal reasoning against external tool use.

To address these challenges, we introduce a comprehensive evaluation framework grounded in the first principles of transformer inference. We propose PTE (Prefill Token Equivalents), a hardware-aware metric that explicitly models the asymmetric costs of the prefill and decode phases. PTE prices the memory-bound decode cost in units of one compute-bound prefill token, giving a single scale that unifies internal reasoning and external tool use while explicitly accounting for non-reusable KV-Cache and long-tool-response scenarios. Therefore, we provide PTE as an efficiency metric that offers an estimate closer to physical hardware behavior in TIR.

To validate our metric, we first demonstrate in a high-concurrency industrial setting that PTE aligns closely with wall-clock latency, significantly outperforming token-count metrics. We further confirm its robustness across various hardware profiles, showing that PTE maintains consistent model efficiency rankings regardless of the specific device.

With this foundation, we conduct extensive experiments across five TIR benchmarks, quantify PTE costs for thousands of trajectories, and dissect four TIR inefficiency patterns: Confirmatory Tool Usage, Tool-Mixing, Lack of Tool Priors, and Tool Format Collapse (as illustrated in Fig. \ref{fig:inefficient_patterns}). Crucially, contrasting PTE with standard token counts reveals a divergence where token generation tends to front-load budgets (Fig. \ref{fig:Step vs token}), whereas PTE exposes how hardware costs actually escalate in later steps due to context accumulation (Fig. \ref{fig:Step vs PTE}). Consequently, we observe that SOTA models with near-identical accuracy can differ by orders of magnitude in PTE. We further identify a clear trade-off in thinking models, finding that while their heavy computational overhead pays off on complex reasoning tasks, it yields diminishing returns on simpler ones. Finally, statistical results indicate that trajectories with higher PTE costs tend to have lower reasoning correctness, suggesting that simply using more tools does not improve answer quality.

The main contributions of our work are summarized as follows:

\begin{compactitem}
    \item We introduce PTE, a hardware-aware metric that unifies reasoning and tool-use costs by explicitly modeling prefill–decode asymmetry in TIR.
    \item We validate PTE in a high-concurrency industrial setting, demonstrating that it aligns significantly better with wall-clock latency than token counts and maintains consistent model rankings across diverse hardware profiles.
    \item Through extensive experiments on five benchmarks, we quantify TIR costs, identify four distinct inefficiency patterns, and reveal a negative correlation between PTE costs and reasoning correctness.
    \item We release a high-concurrency, modular TIR framework featuring flexible tool customization and built-in evaluation to facilitate future research.
\end{compactitem}
\section{Related Work}

\noindent \textbf{Benchmarks for Tool-Augmented LLMs} \quad Early TIR benchmarks focused on API selection and plan decomposition, often abstracting away execution. Examples include BFCL\cite{bfcl}, Webshop\cite{webshop}, ToolBench\cite{toolbench}, and T-Eval\cite{teval}, which primarily measure task success rate. Subsequent datasets like API-Bank\cite{apibank} and Critic-Tool\cite{critictool} introduced execution-based metrics such as API success rate. Recently, Tool-Integrated Reasoning (TIR) benchmarks have evolved to cover complex, multi-step tasks. These include web-browsing (e.g., BrowseComp\cite{browsecomp}, WideSearch\cite{widesearch}, and GAIA\cite{gaia}) and domain-specific reasoning in math and code (e.g., GSM8K\cite{gsm8k}, MATH500\cite{math500}, SWE-Bench\cite{swebench}). Concurrently, Serper\cite{dai2025seper} offers a principled metric for evaluating information-seeking efficiency in tool-augmented systems. While accuracy remains the primary metric, efficiency is often overlooked or measured simply by token or step counts (e.g., MCP-RADAR\cite{mcpradar}, ToolQA\cite{toolqa}, CLASSIC\cite{topoftheclass}). Although recent works attempt to incorporate cost awareness, either through performance-aware cost\cite{dissecting} or economic models for API pricing\cite{economic}, they lack a unified framework grounded in the physical latency of transformer inference.

\noindent \textbf{Efficiency in Tool-Integrated Reasoning} \quad Research has identified inefficient behaviors in TIR, such as "cognitive offloading"\cite{otc} and "over-tooluse"\cite{smart}. To mitigate these, Reinforcement Learning (RL) has been widely adopted, primarily focusing on reward engineering and algorithmic innovation. In terms of reward engineering, approaches fall into two categories: (1) \textbf{Indirect optimization}, which implicitly improves efficiency by supervising reasoning quality, such as evaluating tool variety\cite{toolstar}, the contribution of individual steps\cite{steptool}, or usage correctness\cite{artist}; and (2) \textbf{Direct optimization}, which explicitly incorporates cost penalties. However, current penalties rely on naive metrics like toolcall counts\cite{otc,autotir} or token counts\cite{oppoefficientagent,costbench}, failing to capture hardware-level latency. Beyond reward design, algorithmic innovations aim to optimize the inference process itself. These include entropy-based exploration strategies\cite{ARPO,toollight}, dynamic routing between reasoning and tool-use\cite{a2fm}, and gradient-based stopping criteria\cite{rlev,understanding} to prune redundant steps.
\section{PTE: A First-Principles Efficiency Metric}

\subsection{Background: The Physical Reality of LLM Inference Cost}

\begin{table}[ht]

\begin{threeparttable}
\caption{\textbf{Gamma Values of Different LLMs.} We evaluated a range of state-of-the-art open-source models, all of which were officially declared as having tool-calling capabilities.}
\label{tab1_simple}
\small
\centering
\begin{tabularx}{\linewidth}{l|>{\centering\arraybackslash}X| >{\centering\arraybackslash}X}
\toprule
\textbf{Models} & \textbf{Active Params ($N_{params}$)} & \textbf{Calculated $\gamma$} \\

\midrule
\rowcolor{AliceBlue}
\multicolumn{3}{l}{\textit{\textbf{Open-Source Dense Models}}} \\
Qwen2.5-7B-Instruct & 6.53B & 0.00329 \\
Qwen2.5-32B-Instruct & 31.0B & 0.00320 \\
Qwen2.5-72B-Instruct & 70.0B & 0.00175 \\
Qwen3-32B & 31.2B & 0.00200 \\
Llama-3.1-8B-Instruct & 8.0B & 0.00625 \\
Llama-3.1-70B-Instruct & 70.6B & 0.00175 \\
\midrule
\rowcolor{MistyRose}
\multicolumn{3}{l}{\textit{\textbf{Open-Source MoE Models}}} \\
Qwen3-30B-A3B & 3.3B & 0.00563  \\
Qwen3-235B-A22B-Instruct & 22B & 0.00163 \\
Qwen3-235B-A22B-Thinking & 22B & 0.00163 \\
GLM-4.5-Air & 12B & 0.00200 \\
GLM-4.5 & 32B & 0.00183 \\
DeepSeek-V3.1-Terminus & 37B & 0.00068 \\
GPT-OSS-120B & 5.1B & 0.00388 \\
\bottomrule
\end{tabularx}

\end{threeparttable}
\end{table}

As shown in Fig. \ref{fig:PTEmain}, LLM inference costs are phase-dependent as the prefill and decode phases exhibit different bottlenecks.

\noindent \textbf{Prefill Phase (Compute-Bound)}: Input tokens are processed in parallel, rendering this phase primarily limited by the GPU's computational throughput (FLOPs).

\noindent \textbf{Decode Phase (Memory-Bound)}: Output tokens are generated sequentially. Performance here is bottlenecked by the HBM bandwidth required to load model weights and the Key-Value (KV) Cache. Notably, the cost of retrieving the KV-Cache grows linearly with the cumulative context length ($L_{seq}$).

\noindent \textbf{KV-Cache Eviction}: In the context of TIR, we define eviction broadly to encompass any scenario where the cache state is invalidated during the tool-execution pause, ranging from Time-to-Live (TTL) expiration in continuum serving systems to the architectural necessity of full re-computation in stateless pipelines.


\subsection{Formal Definition of PTE}
To bridge this gap, we introduce PTE (Prefill Token Equivalents), a metric that unifies both compute-bound and memory-bound costs into a single, physically meaningful unit: the equivalent cost of processing one input (prefill) token. For a reasoning trajectory with $k$ turns, the total cost is:

$$ PTE = \sum_{i=1}^{k} (D_{prefill_i} + \gamma \cdot L_{seq_i} \cdot D_{decode_i})$$

Where:
\begin{compactitem}
\item
$D_{prefill_i}$: The total number of context tokens up to turn $i$. This represents the compute-bound prefill cost.
\item
$D_{decode_i}$: The number of tokens generated by the model in turn $i$.
\item
$L_{seq_i}$: The cumulative sequence length (total context) before the decode phase begins in turn $i$.
\item
$\gamma$: A dimensionless coefficient representing the relative cost of a memory-bound operation to a compute-bound one.
\end{compactitem}

\subsection{Cost Modeling}

To quantify the inference cost, we model the computational overhead of the two distinct phases.

\noindent \textbf{Compute-Bound Prefill Cost.} The prefill phase processes input tokens in parallel. Its cost is dominated by matrix multiplications, widely approximated as proportional to the model size\cite{scaling, distserve}:
$$C_{prefill} \approx 2 \cdot N_{params} \text{ [FLOPs]}$$
where $N_{params}$ denotes the number of active parameters.

\noindent \textbf{Memory-Bound Decode Cost.} The decode phase is limited by the memory bandwidth required to load the KV cache. For a model with $n_{layers}$ layers and hidden dimension $d_{model}$ using FP16 precision (2 bytes), the memory access volume is $S_{KV} = 4 \cdot n_{layers} \cdot d_{model} \text{ [Bytes]}$. To unify dimensions with the prefill cost, we convert this memory volume into equivalent computational cost using the Hardware Operational Intensity (HOI)\cite{roofline,rooflinesurvey,efficientdiffusionlanguagemodels}:$$C_{decode}^{eq} = S_{KV} \cdot HOI \text{ [FLOPs]}$$

\noindent \textbf{The $\gamma$ Coefficient.} We define $\gamma$ as the ratio of the equivalent decode cost to the prefill cost, representing the relative penalty of memory-bound operations:
$$\gamma = \frac{C_{decode}^{eq}}{C_{prefill}} = \frac{2 \cdot n_{layers} \cdot d_{model} \cdot HOI}{N_{params}}$$

We refine $\gamma$ to accommodate modern architectural optimizations: for \textbf{Grouped Query Attention (GQA)}, we scale $\gamma$ by the KV-to-Query head ratio ($H_{kv}/H_q$); for \textbf{Multi-Head Latent Attention (MLA)}, we substitute $d_{model}$ with the compressed dimensions ($d_{latent} + d_{rope}$). 

$\gamma$ serves as a static property of a model-hardware pair. Tab. \ref{tab1_simple} lists the results for all evaluated models, while the full statistics are provided in Appendix \ref{sec:gamma_details}. For a deeper interpretation of $\gamma$ as a measure of computational cost scaling efficiency, please refer to Appendix \ref{sec:gamma}.

\begin{figure}[t]
  \centering

  \includegraphics[width=\linewidth]{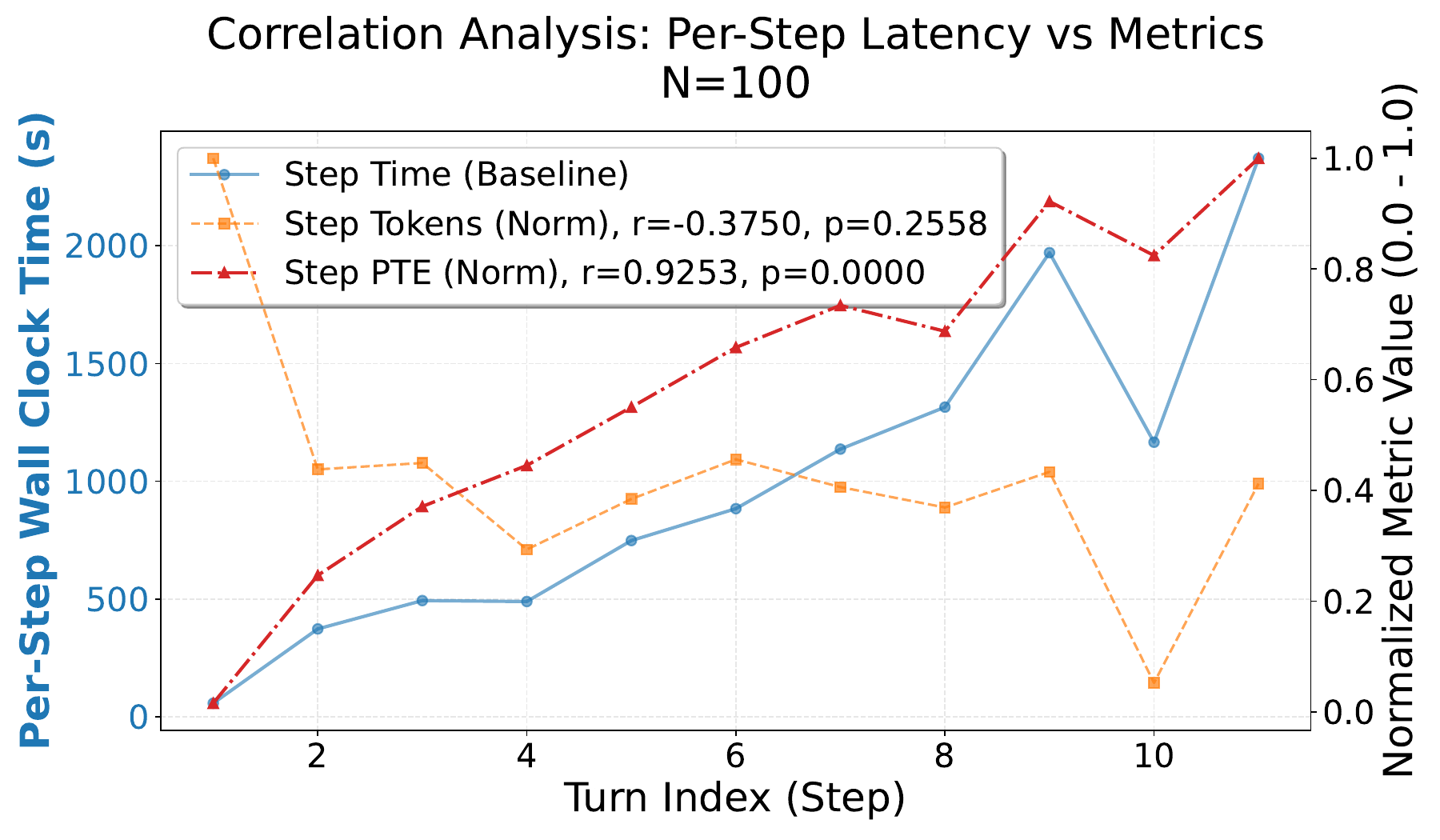}
  
  \caption{Correlation analysis between real-world latency, PTE, and output token counts($N = 100$). Metrics are normalized for visual comparison. PTE demonstrates a strong linear correlation($r = 0.9253$) with wall-clock latency, whereas output token counts show limited correlation($r = -0.3750$).}

  \label{fig:latency}
\end{figure}

\section{Validation: Fidelity and Robustness}
To ensure PTE serves as a reliable efficiency metric, we validate it against two criteria: physical fidelity to real-world latency and robustness across hardware architectures.

\subsection{Validation against Wall-clock Latency}
We conducted high-concurrency experiments using DeepSeek-V3.2 on an $8\times H200$ cluster to simulate industrial TIR scenario. The results in Fig. \ref{fig:latency} reveal a divergence: the raw token counts show limited correlation with actual runtime ($r = -0.3750, N = 100, p = 0.2558$). In contrast, PTE demonstrates a strong positive correlation ($r = 0.9253, N = 100, p < 10^{-4}$) with wall-clock time, significantly outperforming both naive token counts and commercial API pricing metrics. (See Appendix \ref{sec:validation_setup} and \ref{app:api-pricing}). \footnote{Note that step-level statistics are reported for descriptive purposes only, since steps within a trajectory are serially dependent, p-values should not be interpreted as formal hypothesis tests.}

\subsection{Robustness across Hardware Profiles}

\begin{table}[t]
    \centering
    \small
    \setlength{\tabcolsep}{4pt} 
    \caption{\textbf{Robustness of PTE across Hardware Profiles.} High correlations ($\rho > 0.95$) demonstrate that PTE maintains consistent model efficiency rankings despite significant variations in hardware specifications.}
    \label{tab:hardware_robustness}
    \begin{tabular}{lcccc}
    \toprule
    \textbf{Hardware} & \textbf{HOI} & \textbf{$\gamma$ Scaling} & \textbf{Rank Corr.} \\
     & \scriptsize{(FLOPs/Byte)} & \textbf{Factor} \scriptsize{($\alpha$)} & \scriptsize{($\rho$)} \\
    \midrule
    H100 (Ref) & 756.5 & $1.00\times$ & 1.000 \\
    H200       & 348.1 & $0.46\times$ & 0.995 \\
    A100       & 322.5 & $0.43\times$ & 0.989 \\
    RTX 4090   & 327.4 & $0.43\times$ & 0.989 \\
    V100       & 138.9 & $0.18\times$ & 0.956 \\
    \bottomrule
    \end{tabular}
\end{table}

We performed a sensitivity analysis of the coefficient $\gamma$ across devices with varying memory-to-compute ratios. Table \ref{tab:hardware_robustness} summarizes the derived Hardware Operational Intensity (HOI) and the resulting scaling factor $\alpha = \gamma / \gamma_{\text{base}}$ for each device (See Appendix \ref{sec:sensitivity} for detailed hardware specifications). Despite significant variations in hardware characteristics (where $\alpha$ ranges from $0.18\times$ to $1.00\times$), the relative efficiency rankings of models remain highly consistent, with Spearman's rank correlation coefficients ($\rho$) consistently exceeding 0.95. This confirms that PTE captures intrinsic efficiency characteristics independent of the specific deployment platform.
\section{Experimental Setup}
To evaluate TIR efficiency differences using PTE costs, we designed a comprehensive experimental framework covering diverse TIR tasks and models.

\begin{figure*}[t]
  \centering
  \small                         
  \begin{subfigure}[b]{0.48\linewidth}
    \centering
    \includegraphics[width=\linewidth]{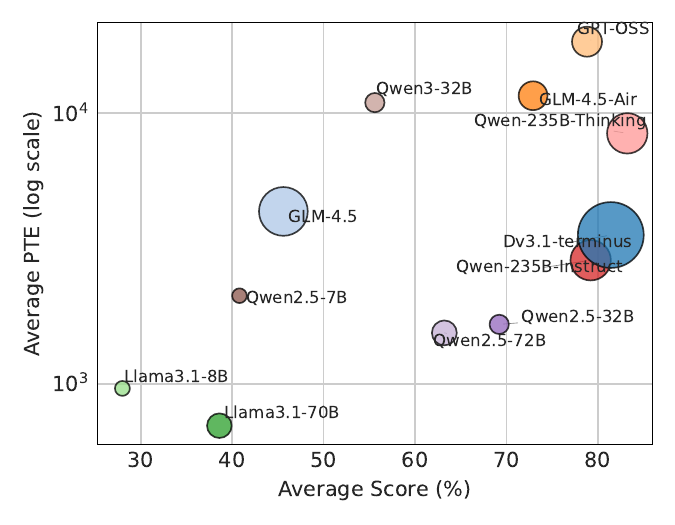}
    \caption{MATH500}
    \label{fig:bench1_1}
  \end{subfigure}
  \hfill                          
  \begin{subfigure}[b]{0.48\linewidth}
    \centering
    \includegraphics[width=\linewidth]{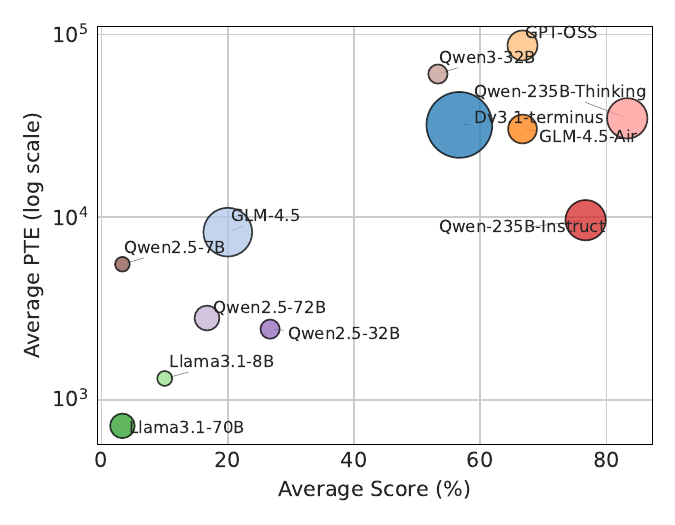}
    \caption{AIME24}
    \label{fig:bench2_1}
  \end{subfigure}

  \begin{subfigure}[b]{0.31\linewidth}
    \centering
    \includegraphics[width=\linewidth]{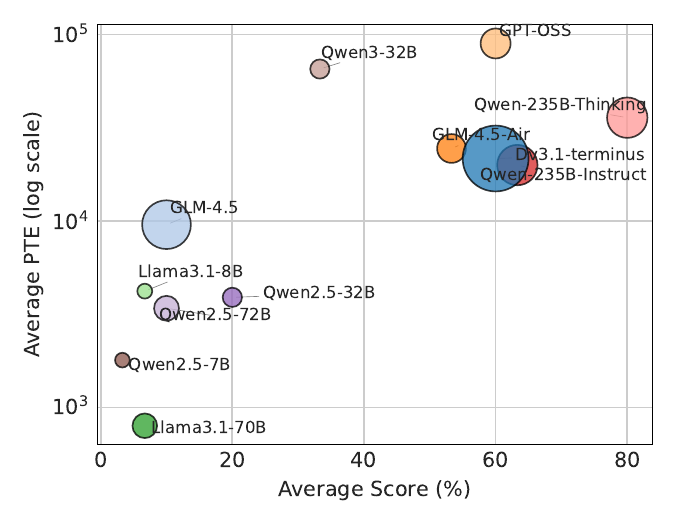}
    \caption{AIME25}
    \label{fig:bench3_1}
  \end{subfigure}
  \hfill
  \begin{subfigure}[b]{0.31\linewidth}
    \centering
    \includegraphics[width=\linewidth]{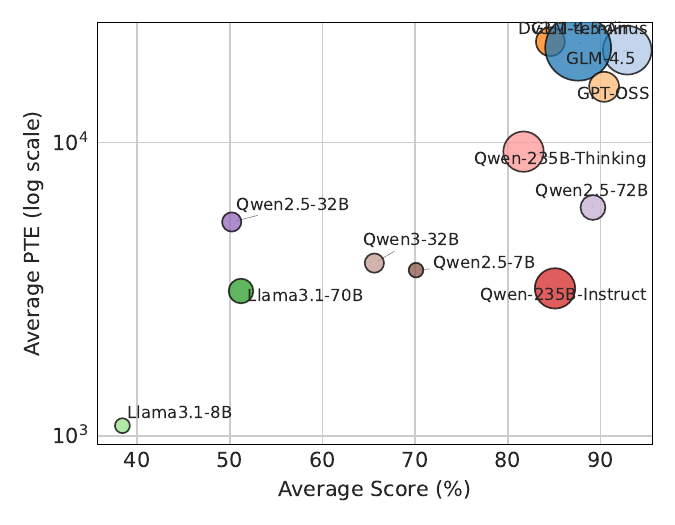}
    \caption{SimpleQA}
    \label{fig:bench4_1}
  \end{subfigure}
  \hfill
  \begin{subfigure}[b]{0.31\linewidth}
    \centering
    \includegraphics[width=\linewidth]{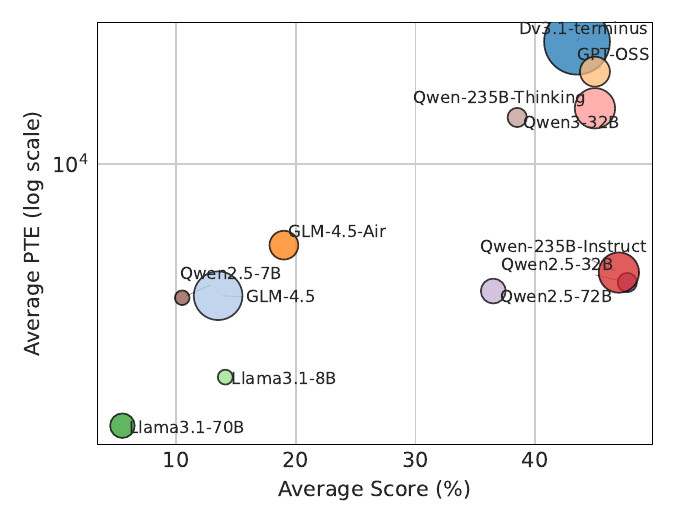}
    \caption{WebInstruct}
    \label{fig:bench5_1}
  \end{subfigure}

  \caption{PTE (\textbf{P}refill-\textbf{T}oken \textbf{E}quivalents) versus Average Score on five benchmarks. The bubble size represents the scale of active parameters. Models in the bottom-right region exhibit better trade-offs between efficiency and accuracy. Note the logarithmic scale on the y-axis.}
  \label{fig:PTE vs score}
\end{figure*}

\subsection{Benchmarks Formulation}

We evaluate models on five benchmarks targeting distinct TIR capabilities. For mathematical reasoning, we use the full test sets of \textbf{MATH500}\cite{math500} (math problems) and \textbf{AIME 2024/2025} (high-difficulty competition problems), equipping agents with a \texttt{Python} tool. For information seeking, we evaluate on randomly sampled subsets of 500 examples from \textbf{SimpleQA}\cite{simpleqa}  (factual QA requiring specific fact retrieval) and \textbf{WebInstruct-Verified}\cite{webinstruct} (complex, multi-disciplinary tasks requiring both retrieval and computation). The former utilizes \texttt{Search} and \texttt{Visit} tools, while the latter adds \texttt{Python} to the toolset. Detailed descriptions of each benchmark are provided in Appendix \ref{sec:benchmarkdetails}.

\subsection{Models and Agent Framework}
We evaluated the tool-capable open-source models listed in Table \ref{tab1} using vLLM. To ensure a fair comparison of intrinsic capabilities, we utilized identical system prompts and tool definitions across all models within our unified framework.

The framework was equipped with three specific tools: (1) a \texttt{Search} tool utilizing Serper API, (2) a \texttt{Visit} tool for webpage content retrieval powered by Jina API, and (3) a \texttt{Python} tool implemented via an open-source Python sandbox.

Our framework orchestrates the iterative TIR process and logs token-level statistics for evaluation. Specifically, at each turn $i$, it records the prefill tokens $D_{prefill_i}$, decoded tokens $D_{decode_i}$, and cumulative sequence length $L_{seq_i}$ (see Appendix \ref{sec:implement} for implementation details).
\section{Results and Analysis}

\subsection{Main Results}
\label{sec:mainresults}

\textbf{Efficiency vs. Accuracy Landscape}. We visualize the main results in Fig. \ref{fig:PTE vs score}, which plots the accuracy of the model against PTE. In this visualization, the bottom-right zone represents better performance in both metrics. More detailed experimental results are available in the Appendix \ref{sec:results}. From the results we conclude three findings about TIR.

\begin{figure*}[t]
  \centering
  \small                         
  \begin{subfigure}[b]{0.48\linewidth}
    \centering
    \includegraphics[width=\linewidth]{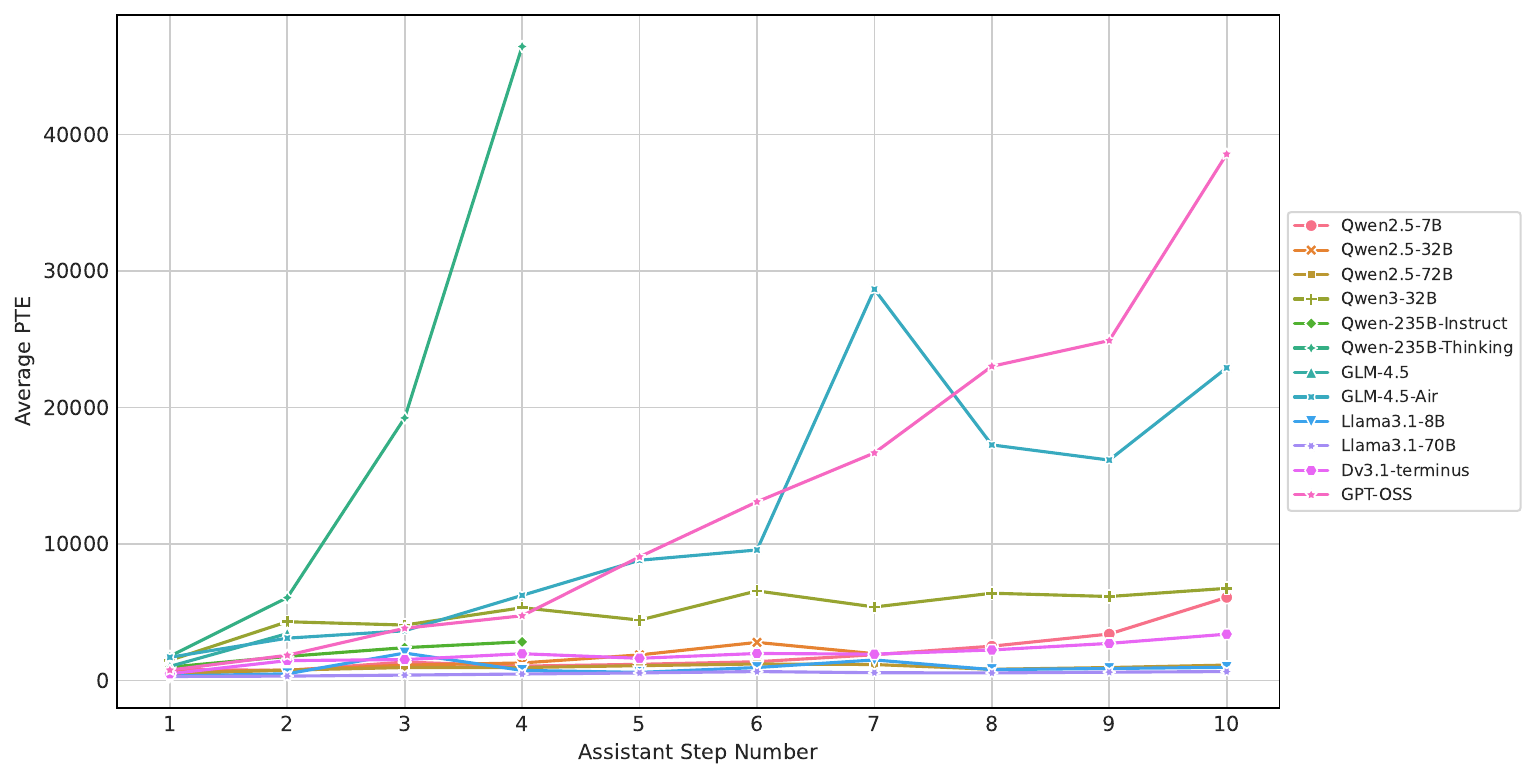}
    \caption{MATH500}
    \label{fig:bench1_3}
  \end{subfigure}
  \hfill                          
  \begin{subfigure}[b]{0.48\linewidth}
    \centering
    \includegraphics[width=\linewidth]{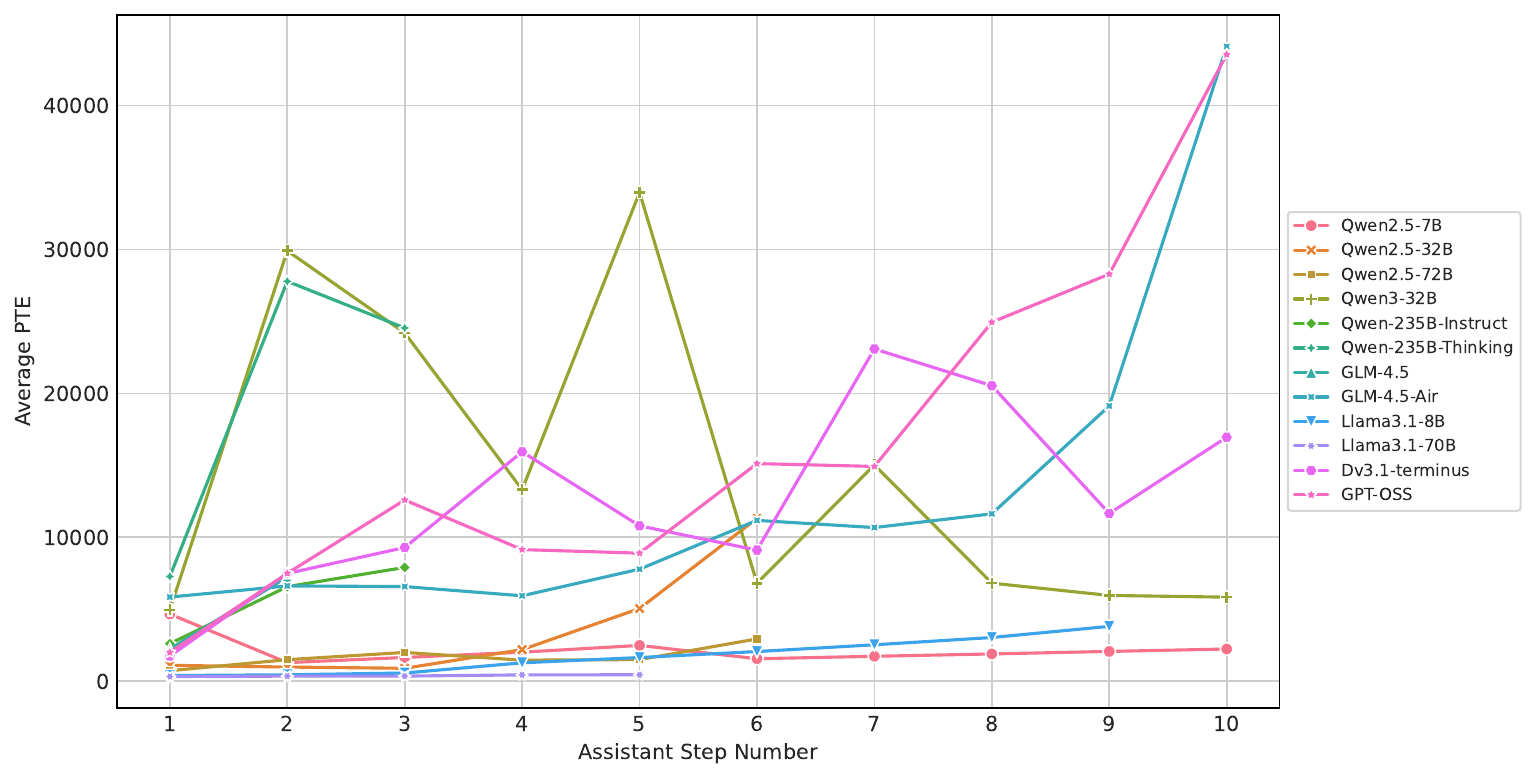}
    \caption{AIME24}
    \label{fig:bench2_3}
  \end{subfigure}

  \begin{subfigure}[b]{0.31\linewidth}
    \centering
    \includegraphics[width=\linewidth]{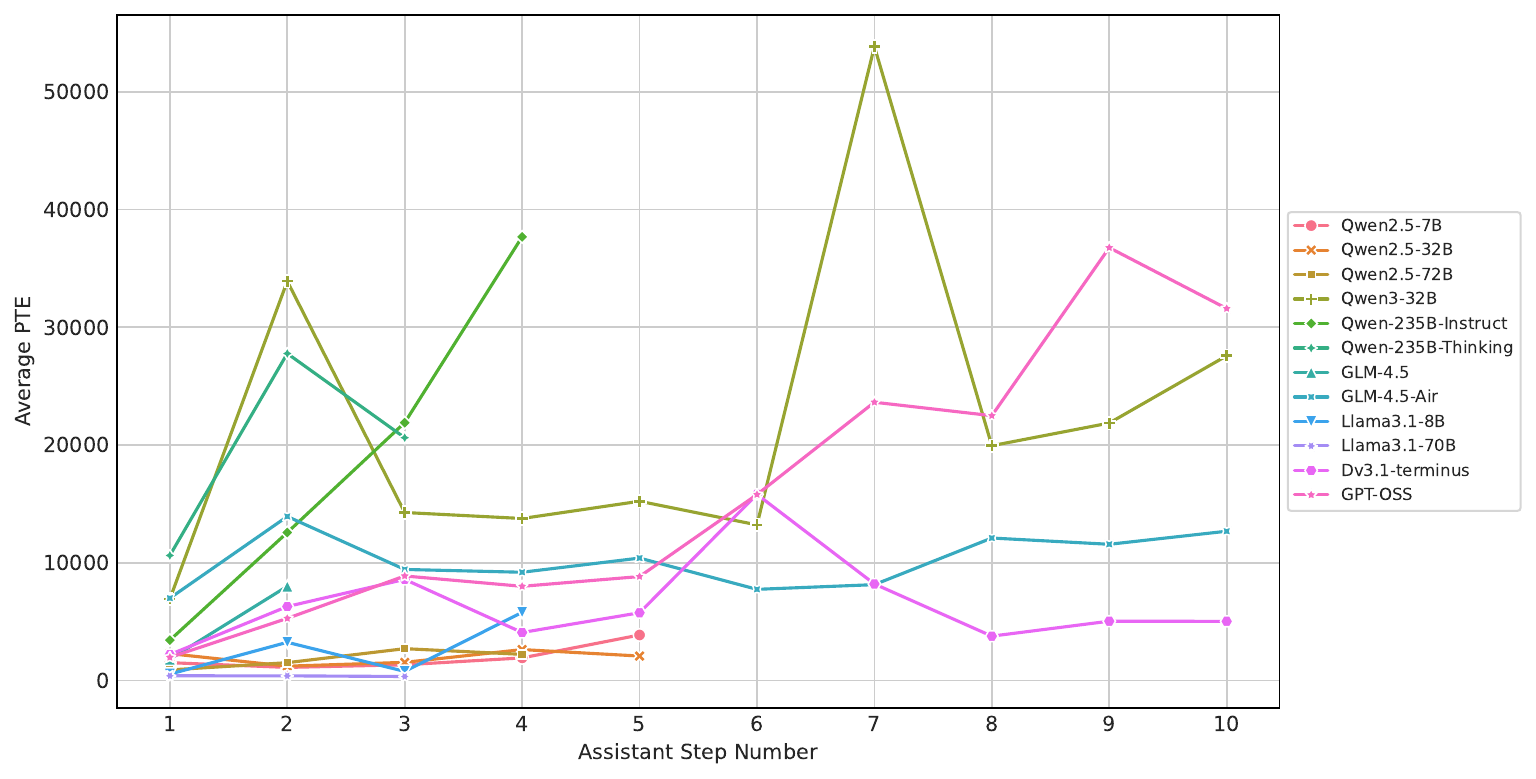}
    \caption{AIME25}
    \label{fig:bench3_3}
  \end{subfigure}
  \hfill
  \begin{subfigure}[b]{0.31\linewidth}
    \centering
    \includegraphics[width=\linewidth]{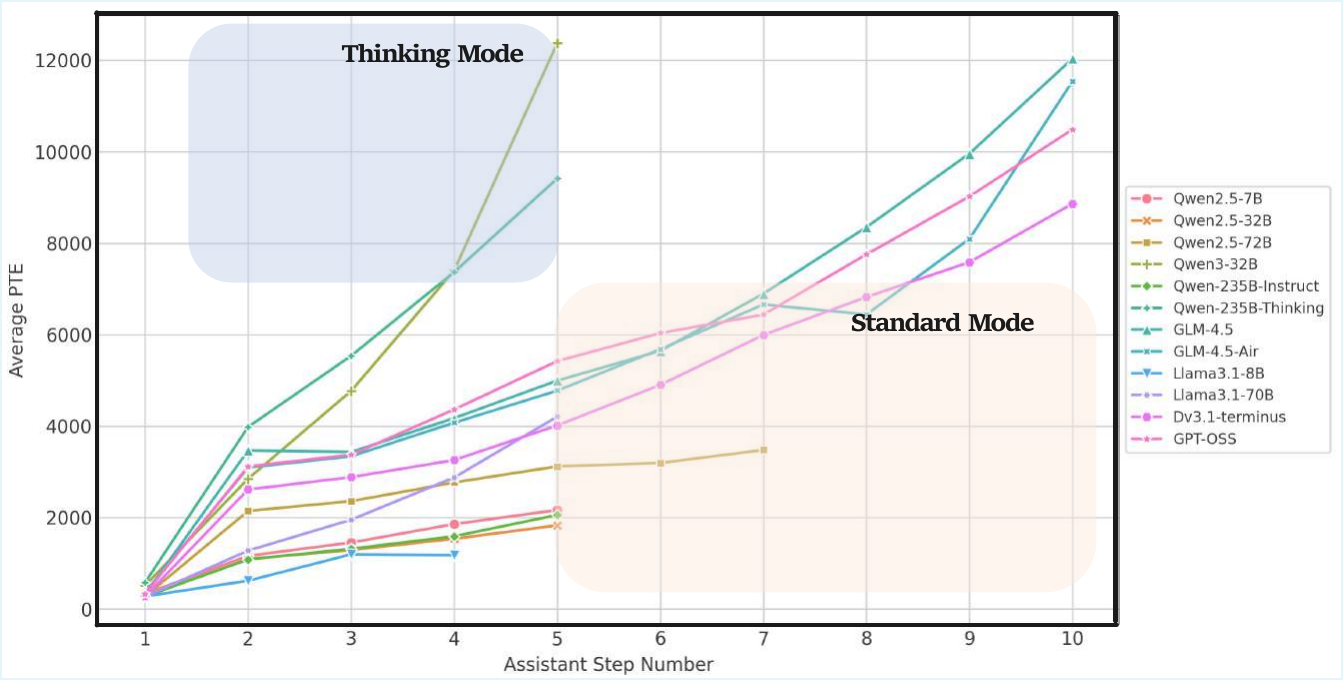}
    \caption{SimpleQA}
    \label{fig:bench4_3}
  \end{subfigure}
  \hfill
  \begin{subfigure}[b]{0.31\linewidth}
    \centering
    \includegraphics[width=\linewidth]{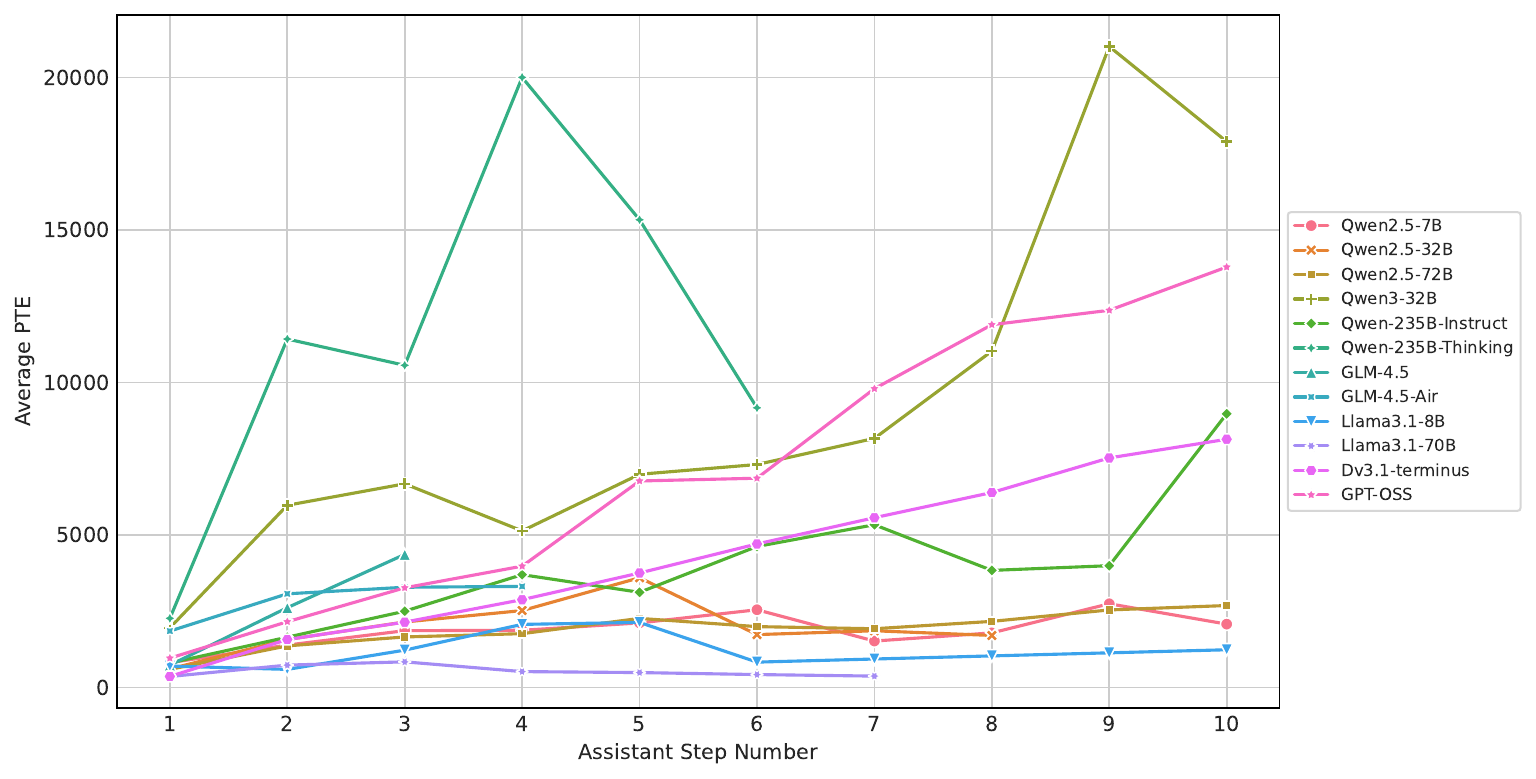}
    \caption{WebInstruct}
    \label{fig:bench5_3}
  \end{subfigure}

  \caption{Distribution of average \textbf{PTE} per reasoning step across five benchmarks. The cost per step escalates as the context length grows, contrasting with the token-based "front-loading" trend. }
  \label{fig:Step vs PTE}
\end{figure*}

\begin{itemize}[leftmargin=*]
    \item \textbf{Vast Cost gap}: Fig. \ref{fig:PTE vs score} shows that while many models cluster at high accuracy (e.g, $\sim70\%$ on AIME24), their corresponding PTE values can span an order of magnitude or more.
    
    \item \textbf{Task-Specific TIR Abilities}: Our data shows that TIR capability is not a general skill but is highly specialized by task and tool type. For example, Qwen2.5-72B excels as a web agent (SimpleQA) but performs poorly as a Python reasoner (MATH500 \& AIME). This suggests that TIR agent evaluations may be task-specific.

    \item \textbf{Different TIR Behavioral Patterns across Models}: We identify five types of different model TIR patterns from the results.
    \begin{itemize}
        \item \textbf{Llama-3.1 series} are pure instruct models. Internal thinking tokens are negligible on every task, yielding high efficiency and medium accuracy.
        \item \textbf{Qwen-2.5 series} achieve high efficiency with moderate accuracy, exhibiting a task-adaptive reasoning strategy. On SimpleQA, the models suppress reasoning to trigger early toolcalls. Conversely, on the four more challenging datasets, they allocate a significant token budget to the initial step, thereby delaying the first tool invocation(See Fig. \ref{fig:Step vs token} in Appendix). We label this phenomenon the “first-step effect” and further analyze it in Sec. \ref{sec:inefficient}.
        \item \textbf{Qwen3-235B-Thinking} and \textbf{Qwen3-32B} (default config) are models with thinking inference mode enabled. They show comparatively high token costs and PTE costs compared to those under standard inference mode (Fig. \ref{fig:bench4_2} and \ref{fig:bench4_3}). On AIME25, the Qwen3-235B-Thinking model yields a +16.7\% accuracy gain over Qwen3-235B-Instruct at only 1.8x PTE cost. However, on the easy SimpleQA task the same model drops 3.4\% in accuracy while PTE grows by 4.2×, indicating severe over-thinking. \textbf{Thinking mode is therefore beneficial only when the task difficulty justifies the extra compute.}
        \item \textbf{Deepseek-V3.1-Terminus}, \textbf{GPT-OSS} and \textbf{Qwen3-235B-Instruct} are frontier models that exhibit relatively high accuracy across most domains. However, the first two are mainly due to their powerful consistent tooluse capacity, whose lengthy and multi-round tool responses noticeably harm efficiency. This suggests that only when agent infrastructures resolve the tooluse KV-Cache eviction problem can these powerful models fully unleash their potential.
        \item \textbf{Tongyi-Deepresearch} is omitted from Fig. \ref{fig:Step vs PTE} and Appendix Fig. \ref{fig:Step vs token} due to recurrent tool format collapse, resulting in an extremely high PTE. This case is discussed in detail in Sec. \ref{sec:inefficient}.
    \end{itemize} 

\end{itemize}

\begin{figure}[t]
  \centering

  \includegraphics[width=\linewidth]{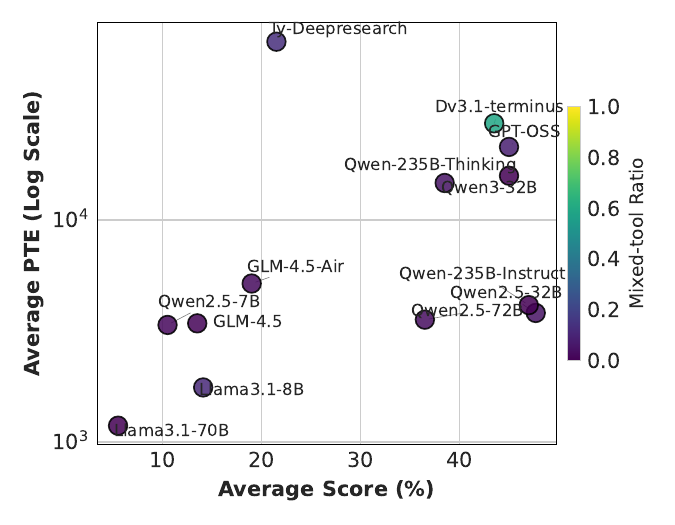}
  
  \caption{Visualization of \textbf{tool-mixing behavior} on WebInstruct-Verified. Point color indicates the ratio of mixed-tool trajectories, with lighter colors (yellow) representing higher mixing frequencies.}

  \label{fig:heatmap}
\end{figure}

\subsection{Analysis of TIR Inefficiency Patterns}
\label{sec:inefficient}
Building on these results, this section provides a deeper analysis of their implications. We evaluate four inefficiency patterns in Tool-Integrated Reasoning through the lens of PTE.

\begin{itemize}[leftmargin=*]
\item \textbf{Confirmatory Tool Usage} Token-count studies on long-horizon reasoning\cite{rhorizon} report a "first-step effect" where models tend to front-load their computational budget, generating the longest trace in the first step. We observe the same token efficiency pattern across most of our TIR models(See Fig. \ref{fig:Step vs token} in Appendix), except for the simplest SimpleQA dataset where reasoning is largely suppressed. This phenomenon largely stems from confirmatory tool usage where models often use tools to verify their internal thoughts rather than using them as direct solvers. Example Trajectory is provided in Appendix \ref{sec:appendix_Z}. 

Our PTE analysis in Figure \ref{fig:Step vs PTE} indicates that a delayed tooluse after a heavy first step significantly increases the accumulated context length($L_{seq}$). This makes every subsequent step more expensive to generate due to the non-reusable KV-Cache and long tool-response in TIR. Consequently, the cost of early overthinking is not immediately obvious, but escalates later as the context grows. The PTE metric highlights this issue and offers a hardware-realistic view of how reasoning budgets are actually consumed in TIR scenarios.

\item \textbf{Tool-mixing} In our investigation into the Webinstruct-Verified benchmark, models were provided with two toolsets: \texttt{search+visit} and \texttt{python}. The heatmap in Fig. \ref{fig:heatmap} reveals that most models rarely mix them, instead commiting to one toolset for the entire trajectory. 

We observed that \textbf{DeepSeek-V3.1-Terminus} was the only model to consistently demonstrate "tool-mixing", a capability that allows it to flexibly combine and alternate between different toolsets within a single reasoning trace. Example Trajectory is provided in Appendix \ref{sec:appendix_A}. However, this impressive tool-mixing capability produces trajectories with significantly higher PTEs without obvious accuracy gain over its peers(See Fig. \ref{fig:bench5_1} and Fig. \ref{fig:heatmap}). We identify tool-mixing phenomenon as a kind of \textbf{inefficient over-tooluse}.

\item \textbf{Lack of Tool Priors} \quad Fig. \ref{fig:PTE vs score} shows that on MATH and AIME the \textbf{Qwen-2.5 series} models tend to lose accuracy and incur higher PTE costs once the \texttt{Python} tool is enabled. We suspect that this reflects limited prior exposure to \texttt{Python} tooluse during pre-training. An example trajectory in Appendix \ref{sec:appendix_B} illustrates how the model often forgets to print the final value, so no output is returned, suggesting that a tool is most efficient when the model has already learned how to use it.

\item \textbf{Tool Format Collapse} \quad {Tongyi-Deepresearch} carries one of the highest PTEs we recorded due to recurrent tool format collapse, as shown in Tab. \ref{tab1_part1} and Tab. \ref{tab1_part2}. It was fine-tuned on a fixed schema and fails when the wording changes even slightly: renaming the tool from “\textbf{search}” to “\textbf{google\_search\_tool}”, or sending a single query instead of the expected list queries, is enough to break the call. These syntactic mismatches lead to significant failures in both accuracy and efficiency. Example trajectories are provided in Appendix \ref{sec:appendix_C}. This is a clear manifestation of losing generalization in TIR, where performance is not robust to non-semantic changes in the input format.

We observe the same tool formats sensitivity in the \textbf{Qwen-2.5 series}. Because it was trained on Python blocks wrapped in triple backticks (\textasciigrave\textasciigrave\textasciigrave python ...\textasciigrave\textasciigrave\textasciigrave), it occasionally reverts to this style instead of using the required <tool\_call> tag. This leads to parsing failures from the vLLM engine.
\end{itemize}

Quantitative breakdowns of these patterns including their detection heuristics, occurrence frequencies, and PTE cost multipliers are detailed in Appendix \ref{app:quantitative-patterns}.
\section{From Efficiency to Intelligence}
\label{sec:effi-acc}

\begin{figure}[t]
  \centering

  \includegraphics[width=\linewidth]{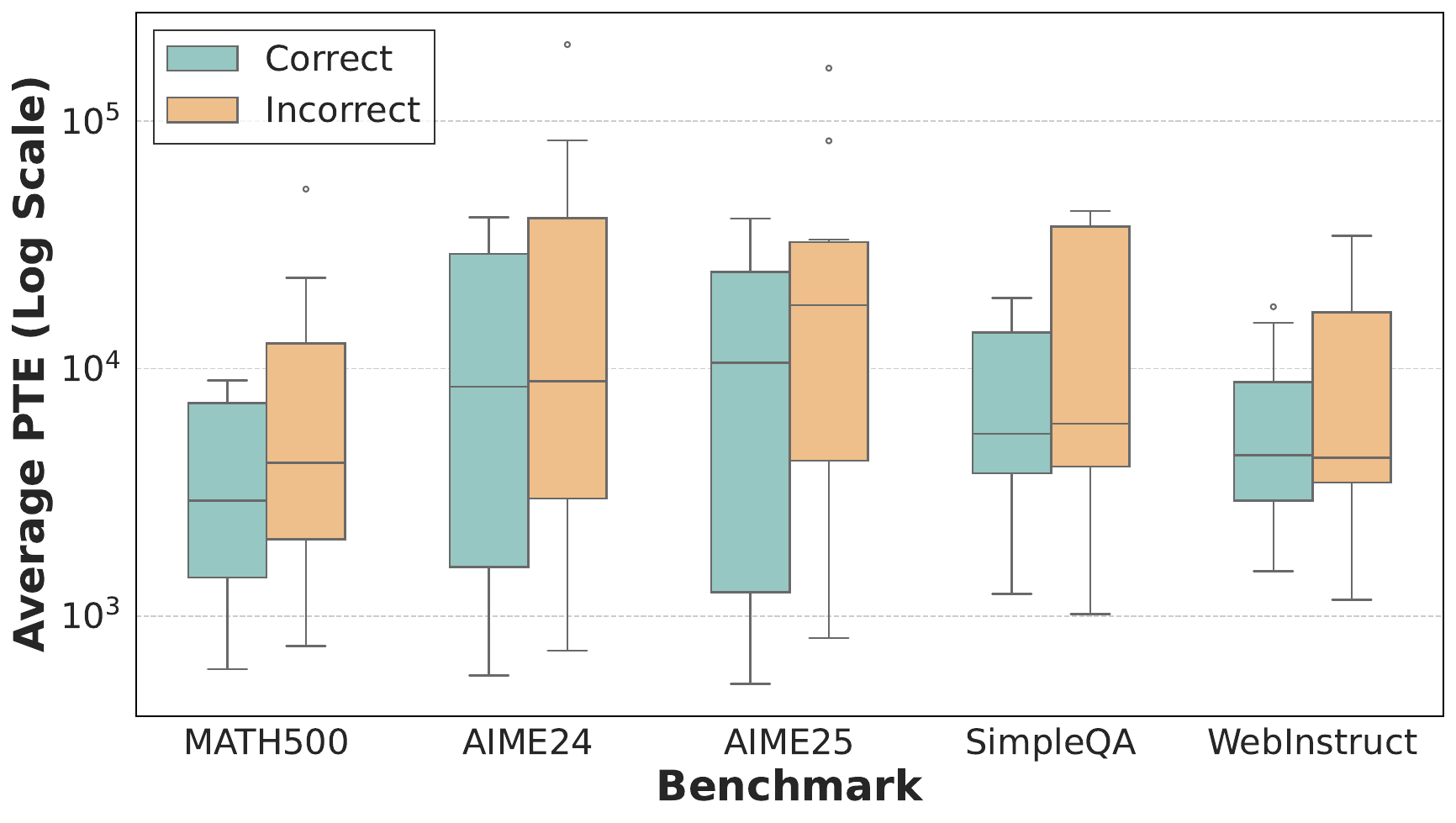}
  
  \caption{Distribution of PTEs for correct and incorrect trajectories across five benchmarks. Incorrect trajectories(right bars) consistently exhibit higher PTE compared to correct ones(left bars). Note the logarithmic scale on the y-axis.}

  \label{fig:score01vsPTE_simplified}
\end{figure}

As shown in Fig. \ref{fig:score01vsPTE_simplified}, trajectories reaching the correct answer consistently exhibit lower PTE values than those finishing with an error across most models and benchmarks. We observe that erroneous trajectories frequently involve prolonged interaction patterns, including repeated toolcalls and extended intermediate reasoning steps, which coincide with elevated PTE cost. Detailed statistical breakdowns for each model and benchmark are provided in Appendix \ref{sec:efficiency_results}, while Appendix~\ref{app:pte-inefficiency} presents difficulty-stratified analyses confirming that these efficiency gaps persist independently of problem hardness.

Qualitative inspection suggests that unsuccessful trajectories often include toolcalls that return unhelpful or irrelevant information, followed by additional steps that attempt to reconcile or reinterpret prior outputs.Such behaviors tend to lengthen the effective sequence length ($L_{seq}$), expand the KV-cache, and are therefore accompanied by higher PTE values. In contrast, many correct trajectories exhibit more compact interaction patterns, with fewer detours and shorter contexts.

Importantly, these observations are correlational. Rather than implying that higher PTE causes errors, our results indicate that elevated PTE frequently co-occurs with reasoning processes marked by uncertainty, redundancy, or ineffective tool utilization. From this perspective, PTE may serve as a coarse-grained diagnostic signal for identifying potentially inefficient reasoning trajectories.
\section{Conclusion}
We introduce PTE, a hardware-aware metric unifying reasoning and tool-use costs by modeling the prefill-decode asymmetry. Experiments across models and tasks reveal that TIR costs span orders of magnitude, largely driven by four specific inefficiency patterns. We also find a correlation between low PTE and high accuracy of trajectory. Hopefully, these findings may provide a unified view of TIR efficiency and offer a new perspective for future work in this field.
\clearpage 

\section{Limitations}

Several limitations exist. Firstly, PTE measures transformer computation, omitting real-world costs like API latency. Secondly, the $\gamma$ parameter represents a simplified abstraction of architectural efficiency; while useful, it may not fully capture all nuances of specific hardware optimizations or runtime dynamics. Finally, our empirical validation was limited to specific tasks and models; the generalizability of our findings, particularly the link between low PTE and high-quality reasoning, requires further exploration across broader domains and architectures.

\section{Ethical Considerations}
This work introduces a new metric to evaluate the efficiency of Large Language Models in tool-integrated reasoning scenarios. We believe that our proposed metric, PTE, contributes positively to the community by enabling researchers and practitioners to identify computational inefficiencies, potentially leading to reduced energy consumption and a lower carbon footprint for deploying LLM agents.

The datasets used in our experiments (MATH500, AIME, SimpleQA, etc.) are all publicly available and standard in the field. We adhered to the terms of use for all APIs and models employed during our evaluation. We do not foresee any immediate negative ethical consequences or societal harm resulting from the proposed metric or the analysis of inefficiency patterns.

\section*{Acknowledgments}
This work was supported by the Anhui Provincial Natural Science Foundation under Grant 2108085UD12. We acknowledge the support of GPU cluster built by MCC Lab of Information Science and Technology Institution, USTC. The AI-driven experiments, simulations and model training were performed on the robotic AI-Scientist platform of Chinese Academy of Sciences.
\bibliography{custom}
\appendix
\clearpage 

\section{Metric Formulation and Hardware Grounding}

\subsection{Model Architectural Details}
\label{sec:gamma_details} 
Due to space constraints, the main text presents a simplified view of the model specifications. Tab. \ref{tab1} provides the complete architectural details for all evaluated models, including the number of layers ($n_{layers}$), hidden dimensions ($d_{model}$), and the architectural factor ($H_{kv}/H_{q}$) used to calculate the $\gamma$ coefficient.

\begin{table*}[ht]

\begin{threeparttable}
\caption{\textbf{Gamma Values of Different LLMs.} We evaluated a range of state-of-the-art open-source models, all of which were officially declared as having tool-calling capabilities.}
\label{tab1}
\small
\centering
\begin{tabularx}{\textwidth}{l|>{\centering\arraybackslash}X >{\centering\arraybackslash}X >{\centering\arraybackslash}X >{\centering\arraybackslash}X| >{\centering\arraybackslash}X}
\toprule
\textbf{Models} & \textbf{Active Params ($N_{params}$)} & \textbf{Layers ($n_{layers}$)} & \textbf{Hidden Dim ($d_{model}$)} & \textbf{Architecture Factor ($H_{kv}/H_q$)} & \textbf{Calculated $\gamma$} \\

\midrule
\rowcolor{AliceBlue}
\multicolumn{6}{l}{\textit{\textbf{Open-Source Dense Models}}} \\
Qwen2.5-7B-Instruct & 6.53B & 28 & 3584 & $1/7$ & 0.00329 \\
Qwen2.5-32B-Instruct & 31.0B & 64 & 5120 & $1/2$ & 0.00320 \\
Qwen2.5-72B-Instruct & 70.0B & 80 & 8192 & $1/8$ & 0.00175 \\
Qwen3-32B & 31.2B & 64 & 5120 & $1/8$ & 0.00200 \\
Llama-3.1-8B-Instruct & 8.0B & 32 & 4096 & $1/4$ & 0.00625 \\
Llama-3.1-70B-Instruct & 70.6B & 80 & 8192 & $1/8$ & 0.00175 \\
\midrule
\rowcolor{MistyRose}
\multicolumn{6}{l}{\textit{\textbf{Open-Source MoE Models}}} \\
Qwen3-30B-A3B & 3.3B & 48 & 2048 & $1/8$ & 0.00563  \\
Qwen3-235B-A22B-Instruct & 22B & 94 & 4096 & $1/16$ & 0.00163 \\
Qwen3-235B-A22B-Thinking & 22B & 94 & 4096 & $1/16$ & 0.00163 \\
GLM-4.5-Air & 12B & 46 & 4096 & $1/12$ & 0.00200 \\
GLM-4.5 & 32B & 92 & 5120 & $1/12$ & 0.00183 \\
DeepSeek-V3.1-Terminus & 37B & 61 & 576 & -- & 0.00068 \\
GPT-OSS-120B & 5.1B & 36 & 2880 & $1/8$ & 0.00388 \\
\bottomrule
\end{tabularx}

\begin{tablenotes}
  \small 
  \item[1] All architectural parameters are sourced from the models' public config.json files.
  \item[2] Llama 3.1 series does not report active parameters. We assume active parameters equals to total parameters.
  \item[3] For DeepSeek-V3, the $d_{model}$ is substituted by $d_{latent} + d_{rope}$, reflecting the significant KV-Cache optimization achieved by its Multi-Head Latent Attention (MLA) architecture.
  \item[3] We use the NVIDIA H100 PCIe as our reference hardware for consistency, which has a Hardware Operational Intensity (HOI) of approximately 756.5 FLOPs/Byte, as detailed in Appendix \ref{sec:HOI}.
\end{tablenotes}

\end{threeparttable}
\end{table*}

\begin{figure*}[t]
  \centering
  \small                         
  \begin{subfigure}[b]{0.48\linewidth}
    \centering
    \includegraphics[width=\linewidth]{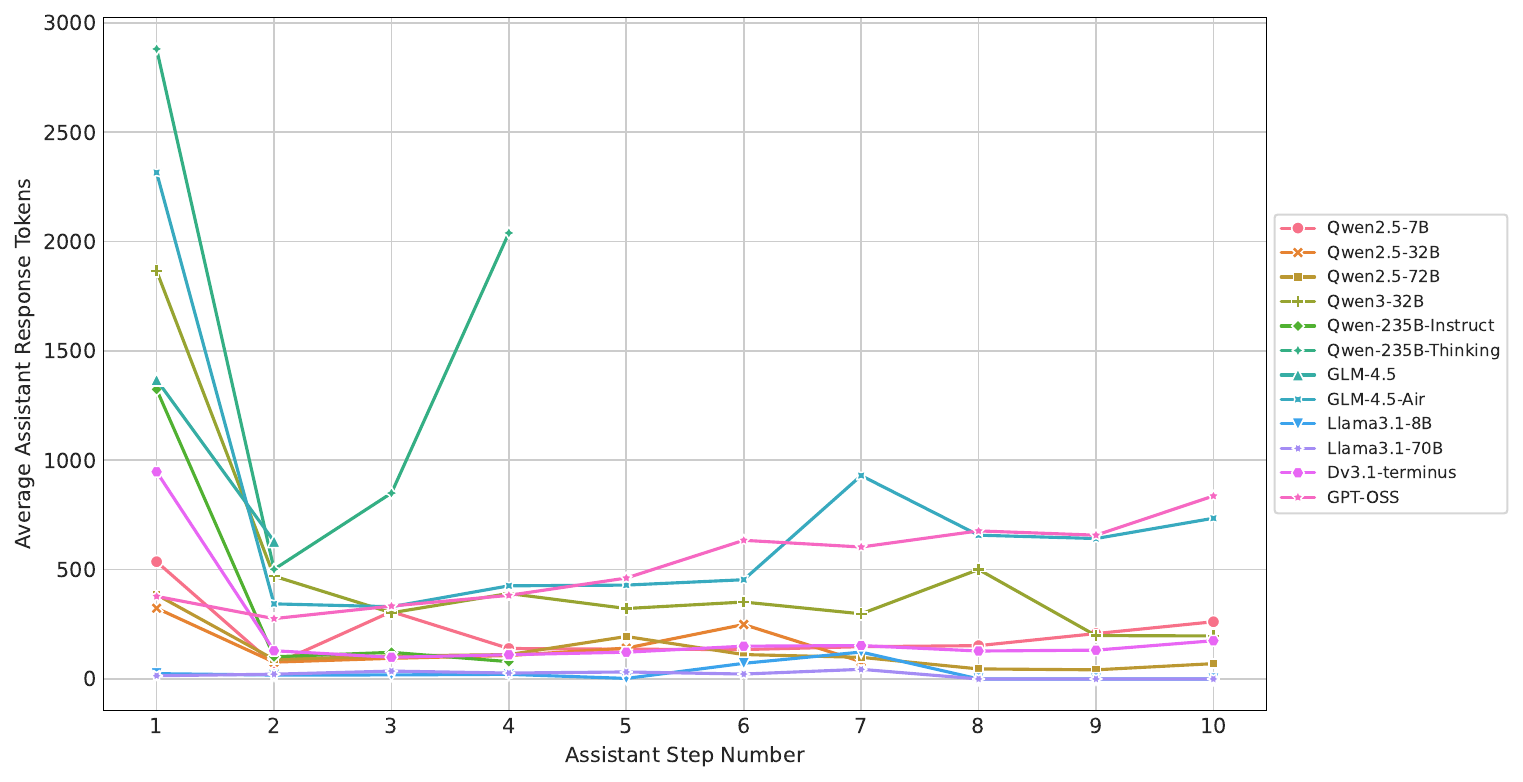}
    \caption{MATH500}
    \label{fig:bench1_2}
  \end{subfigure}
  \hfill                          
  \begin{subfigure}[b]{0.48\linewidth}
    \centering
    \includegraphics[width=\linewidth]{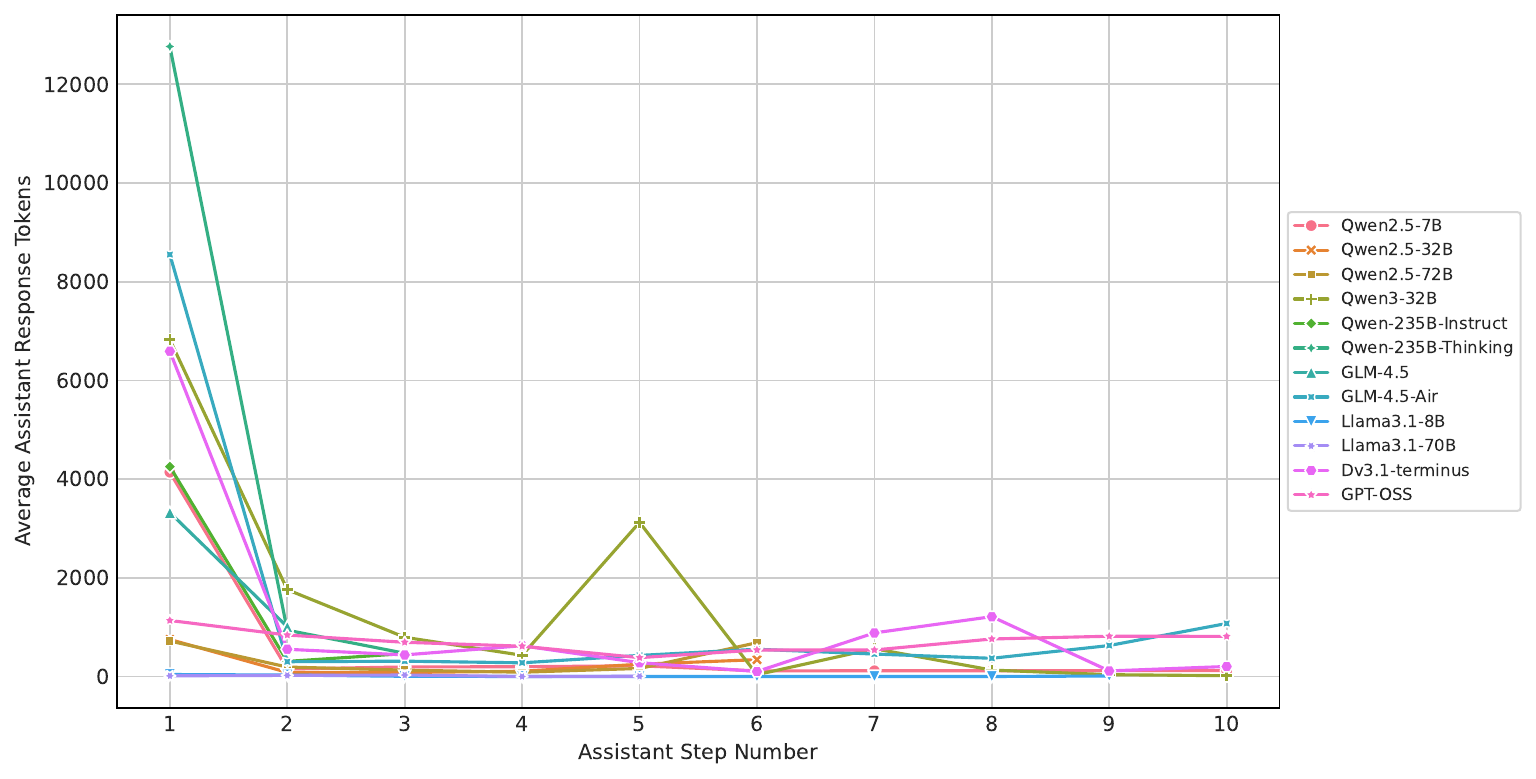}
    \caption{AIME24}
    \label{fig:bench2_2}
  \end{subfigure}

  \begin{subfigure}[b]{0.31\linewidth}
    \centering
    \includegraphics[width=\linewidth]{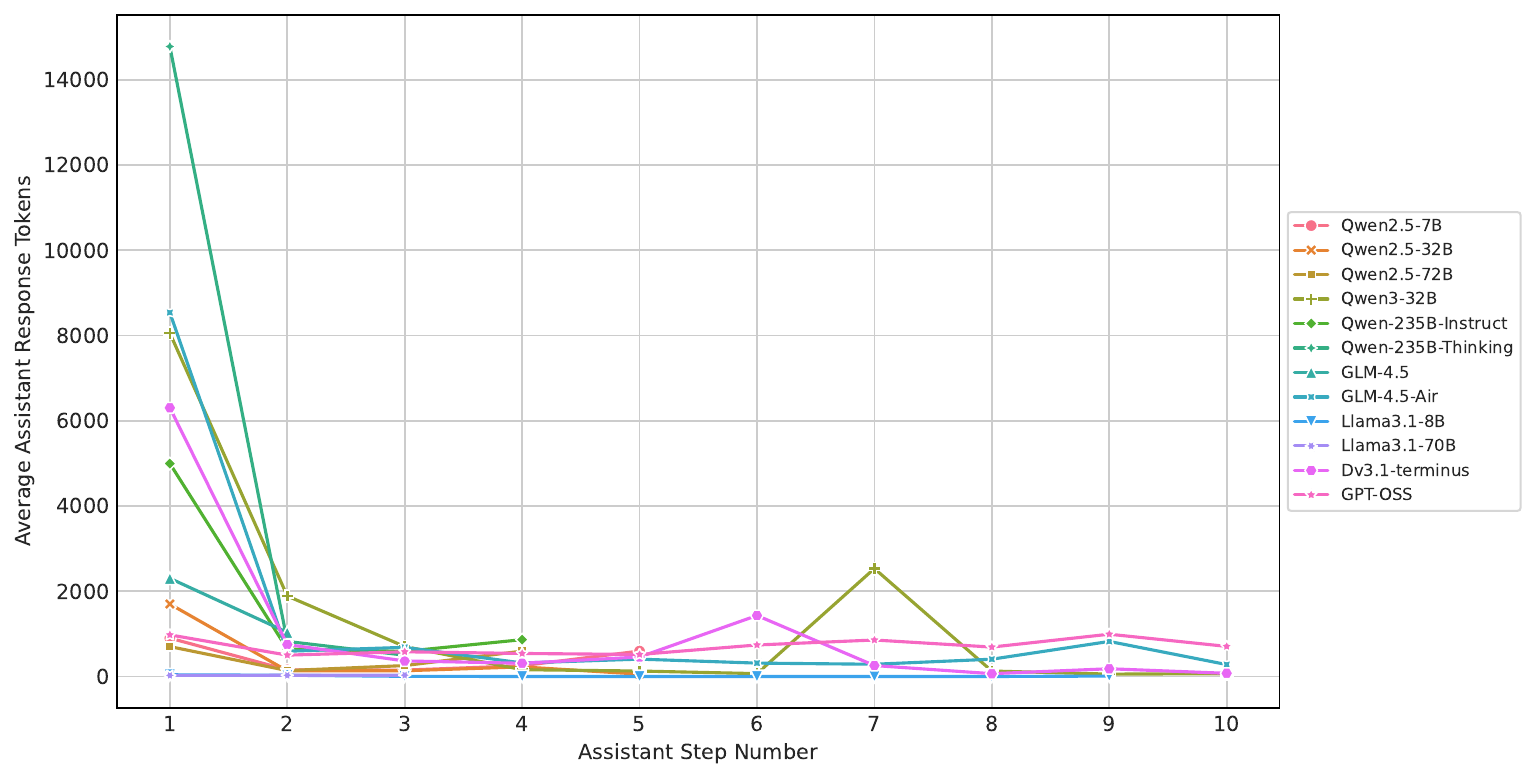}
    \caption{AIME25}
    \label{fig:bench3_2}
  \end{subfigure}
  \hfill
  \begin{subfigure}[b]{0.31\linewidth}
    \centering
    \includegraphics[width=\linewidth]{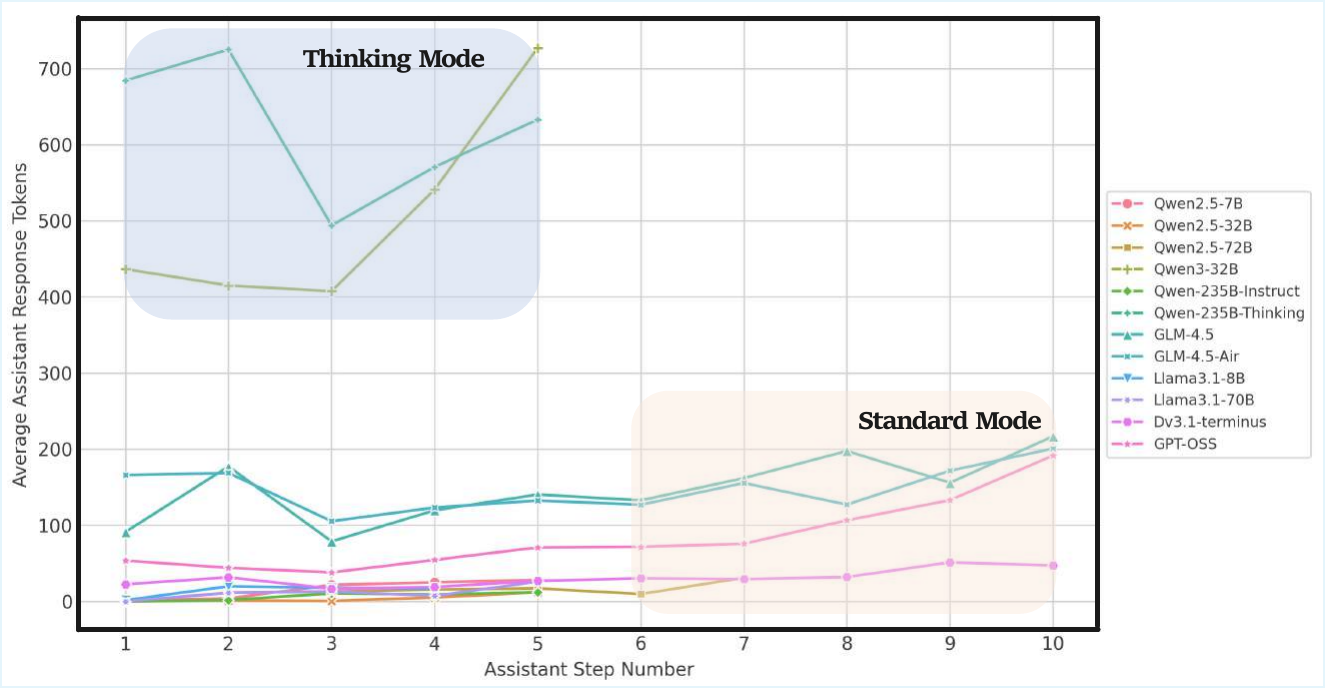}
    \caption{SimpleQA}
    \label{fig:bench4_2}
  \end{subfigure}
  \hfill
  \begin{subfigure}[b]{0.31\linewidth}
    \centering
    \includegraphics[width=\linewidth]{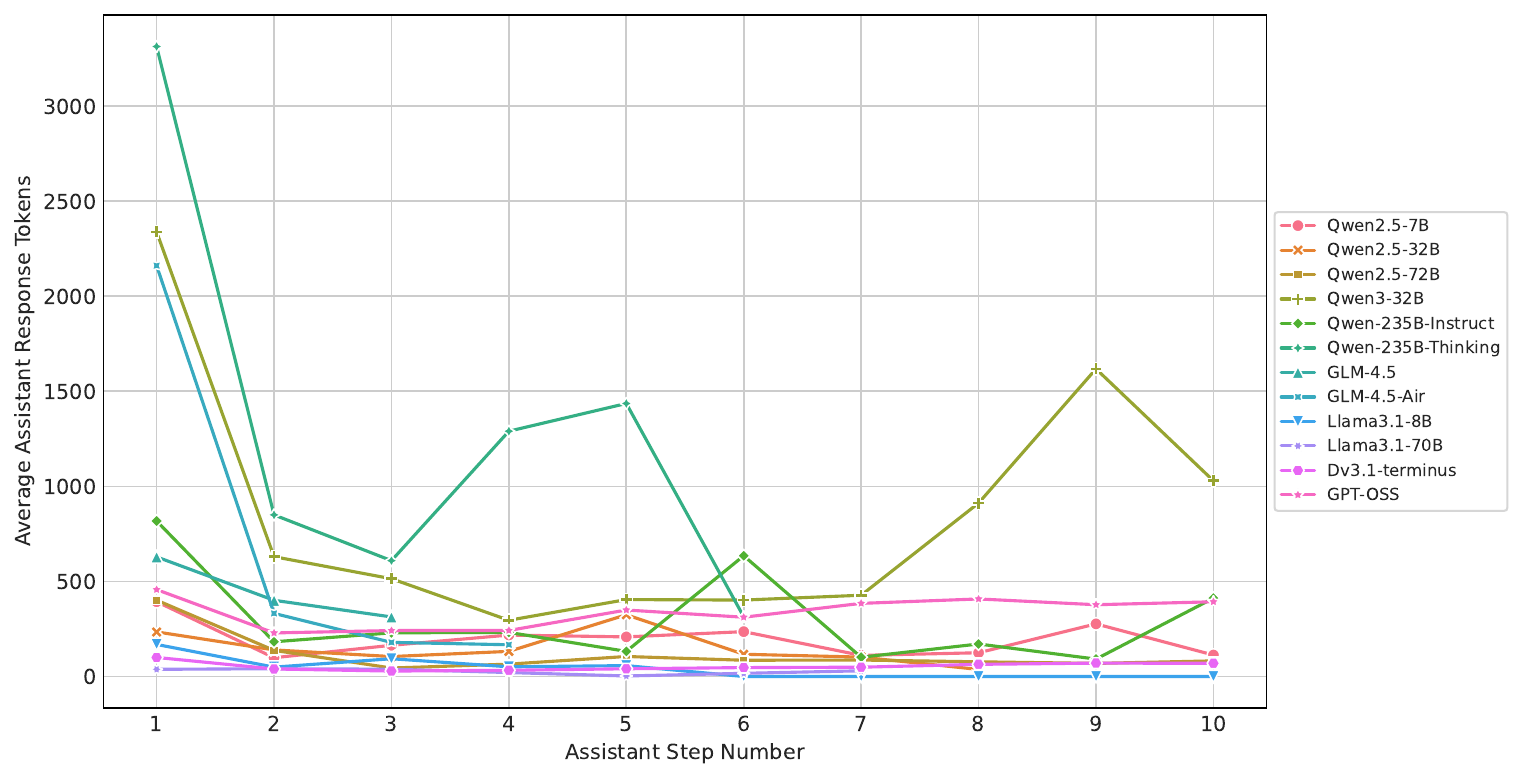}
    \caption{Webinstruct}
    \label{fig:bench5_2}
  \end{subfigure}

  \caption{Distribution of Average Assistamt \textbf{Response Tokens} Across Reasoning Steps. The figure illustrates the response length for each step in a reasoning trajectory. The phenomenon is described as "\textbf{first-step effect}" where models tend to "front-load" their computational budget.}
  \label{fig:Step vs token}
\end{figure*}

\subsection{Hardware Operational Intensity Details}
\label{sec:HOI}
To translate memory size into a computational cost, we employ an important concept from the field of high-performance computing (HPC): Hardware Operational Intensity (HOI)\cite{roofline,rooflinesurvey,efficientdiffusionlanguagemodels}. HOI is an architectural constant that defines the ridge point in the hardware Roofline model. It represents the ratio between the hardware's peak computational throughput (FLOPs/s) and its peak memory bandwidth (Bytes/s).

$$\text{HOI} = \frac{\text{Peak FLOPs}}{\text{Peak Memory Bandwidth}}$$

For consistency, we utilize the NVIDIA H100 80GB PCIe (Hopper architecture) as our reference hardware. These numerator and denominator values are architectural constants provided in official vendor documentation, such as the \textbf{NVIDIA H100 Tensor Core GPU Datasheet}.

\textbf{Peak Memory Bandwidth}: The official specifications for the H100 80GB PCIe (HBM2e) version indicate a peak memory bandwidth of $2.0 \text{ TB/s}$ (i.e., $2.0 \times 10^{12} \text{ Bytes/s}$).

\textbf{Peak Computational Performance}: For the FP16/BF16 precision, the H100 specification sheet indicates a peak theoretical throughput of approximately $1,513 \text{ TFLOPS}$, a figure that accounts for architectural features such as the Transformer Engine. As the Roofline model's definition of HOI relies on the architecture's theoretical performance ceiling, we utilize this peak value (i.e., $1,513 \times 10^{12} \text{ FLOPs/s}$) for the numerator.

\textbf{HOI Calculation}: 
\begin{equation}
\begin{aligned}
\text{HOI}
&= \frac{1{,}513 \times 10^{12}~\text{FLOPs/s}}{2.0 \times 10^{12}~\text{Bytes/s}} \\[4pt]
&= \mathbf{756.5~\text{FLOPs/Byte}}
\end{aligned}
\end{equation}

Thus, we use the value $756.5$ to represent the theoretical peak operational intensity for the H100 architecture at FP16/BF16 precision. We utilize this architectural constant to anchor our $\gamma$ coefficient.

\subsection{Sensitivity Analysis of $\gamma$ Across Hardware Architectures}
\label{sec:sensitivity}
To assess the robustness of the PTE metric across different hardware environments, we conducted a sensitivity analysis of the parameter $\gamma$. As defined before, $\gamma$ is linearly proportional to the Hardware Operational Intensity ($HOI$):

$$\gamma \propto HOI= \frac{\text{Peak FLOPs}}{\text{Peak Memory Bandwidth}}$$

We define a scaling factor $\alpha = \gamma / \gamma_{\text{base}}$, using the NVIDIA H100 as the baseline. Tab. \ref{tab:hardware_comparison} presents the comparative analysis across various architectures, including the H200, A100, V100, and RTX 4090. The four devices are selected to give the sweep a footprint as wide as possible in the real-world hardware landscape. The variation in $\alpha$ reflects the evolving battle against the "Memory Wall." However, the newer H200 significantly reduces $\gamma$ ($0.46\times$) by more than doubling the memory bandwidth, demonstrating specific hardware optimizations targeting memory bottlenecks. 

To verify the stability of our efficiency metric under these varying conditions, we re-evaluated the model rankings on the WebInstruct-Verified dataset trajectories by simulating PTE costs across different hardware profiles. Despite significant variations in hardware characteristics (where $\alpha$ ranges from $0.18\times$ to $1.00\times$), the relative efficiency rankings of the models remain highly consistent. This is evidenced by the Spearman's rank correlation coefficients ($\rho$) reported in Tab. \ref{tab:hardware_comparison}, which consistently exceed 0.95 across all devices.

In summary, while absolute PTE costs fluctuate with hardware specifications, the PTE metric can still capture some intrinsic efficiency characteristics of TIR models, indicating its potential to be an indicator independent of the specific deployment platform.

\begin{figure*}[!ht] 
\begin{tcolorbox}[title = {System Prompt For Webinstruct-Verified}]
You are an \textbf{Expert Problem Solver}. Your mission is to accurately solve problems by skillfully integrating reasoning with your available tools: \texttt{python\_tool}, \texttt{google\_search\_tool}, and \texttt{visit\_tool}. As you proceed, adhere to the following principles:

1.  \textbf{Strategic Tool Selection}: For math/data, you MUST use \texttt{python\_tool}. For facts/real-world info, you MUST use \texttt{google\_search\_tool} and \texttt{visit\_tool}. For complex problems, use web tools to gather data first, then \texttt{python\_tool} to compute.

2.  \textbf{Rigorous Verification}: Before giving a Final Answer, you MUST use \texttt{python\_tool} to verify all calculations and use your web tools to cross-check all information for accuracy.

3.  \textbf{Effective Tool Use}: Write clean, executable Python code, remembering to use \texttt{print()} for output. Formulate precise search queries for \texttt{google\_search\_tool} and specify a clear goal for \texttt{visit\_tool}.

4.  \textbf{Persistence}: Do not give up. Engage in as many tool calls as necessary to thoroughly explore the topic and find a complete, accurate answer.

5.  \textbf{Final Answer Format}: After completing all reasoning and verification, provide the final answer and nothing else, wrapped in \texttt{<ANSWER>answer</ANSWER>}.
\end{tcolorbox}
\caption{System Prompt Example}
\label{fig:sysprompt}
\end{figure*}

\begin{figure*}[!ht] 
\begin{tcolorbox}[title = {System Prompt For LLM Judge}]
Predicted Answer: \texttt{predicted answer}

Expected Answer: \texttt{ground truth}

Determine if the predicted answer is semantically correct (case-insensitive). 

Reply \textbf{only} a JSON object in the format: \{"correct": 0\} or \{"correct": 1\} 

Do not include any additional text or markdown formatting.
\end{tcolorbox}
\caption{Answer Verify Prompt}
\label{fig:llmjudge}
\end{figure*}

\subsection{Interpreting \texorpdfstring{$\gamma$}{gamma}: Computational Cost Scaling Efficiency}
\label{sec:gamma}
Our analysis suggests that $\gamma$ should be interpreted as a measure of computational cost scaling efficiency rather than absolute per-token latency. A low $\gamma$ may not indicate a faster model in terms of tokens-per-second, but rather that the model's total infer cost grows more slowly as the sequence length ($L_{seq}$) increases. This architectural trade-off is more evident when comparing different Mixture-of-Experts (MoE) models from Tab. \ref{tab1}.

For instance, \textbf{GPT-OSS-120B} exhibits a relatively high $\gamma$ ($\approx 0.00388$), which seems to be correlated with its computation-light architecture (5.1B activated parameters). This design appears to be more sensitive to the memory overhead of the KV cache, leading to less efficient scaling. Conversely, a computation-heavy model such as \textbf{DeepSeek-V3.1-Terminus} (37B activated parameters) presents a much lower $\gamma$ ($\approx 0.00068$), suggesting its cost scales more slowly. This implies that its massive computational cost significantly outweighs its memory cost at longer context lengths. This comparison indicates that $\gamma$ may be a useful metric to identify architectures that are more efficient in non-reusable KV-Cache and long-context scenarios.

\section{Experimental Setup}

\subsection{Benchmark Details}
\label{sec:benchmarkdetails}
Our benchmark suite was selected to assess LLMs' TIR capabilities, requiring both internal reasoning and external tool use. Recent surveys on inductive reasoning abilities of LLMs \cite{chen2025surveyinductivereasoninglarge} provide comprehensive taxonomies of reasoning patterns that underpin these benchmark designs. We evaluated the common LLMs on five distinct benchmarks, each targeting different capability aspects:

\begin{itemize}[leftmargin=*]
    \item \textbf{MATH500}\cite{math500}: A benchmark focused on \textbf{mathematical reasoning}, consisting of a collection of math word problems that require numerical computation. We provide \texttt{Python} tool for this benchmark.
    \item \textbf{AIME24, AIME25}: A \textbf{competition math} benchmark featuring high-difficulty problems from the American Invitational Mathematics Examination.We provide \texttt{Python} tool for these benchmarks.
    \item \textbf{SimpleQA}\cite{simpleqa}: A benchmark for \textbf{factual question answering}. It contains questions that require the retrieval of specific facts from external knowledge sources. We provide \texttt{Search} and \texttt{Visit} tools for this benchmark. For our evaluation, we use a randomly sampled subset of 500 examples to ensure efficient yet representative testing.
    \item \textbf{Webinstruct-Verified}\cite{webinstruct}: A complex, \textbf{multi-disciplinary QA} benchmark. It is designed to test models on tasks that may require a combination of information retrieval and computation to answer. We provide \texttt{Python} and \texttt{Search} and \texttt{Visit} tools for this benchmark. For our evaluation, we use a randomly sampled subset of 500 examples to ensure efficient yet representative testing.
\end{itemize}

\subsection{Implementation Details}
\label{sec:implement}
We used the paid Serper and Jina APIs for the \texttt{search} and \texttt{visit} tools, and adopted the open-source SandboxFusion for \texttt{python} tool execution. All tool-call generation and parsing were handled by the vLLM inference engine, on which every tested model was deployed. 

\subsection{Latency Validation Setup}
\label{sec:validation_setup}
To rigorously evaluate the correlation between the proposed PTE metric and real-world latency, We conducted a high-throughput evaluation experiment to mimic industrial Tool-Integrated Reasoning (TIR) workloads. We deployed DeepSeek-V3.2 on an 8$\times$H200 GPU node via vLLM (TP=8), simulating a high-concurrency scenario with 256 parallel requests. The evaluation employed a synthetic dataset of complex tool-requiring QAs derived from Wikidata, requiring the model to utilize \texttt{Search}, \texttt{Visit}, and \texttt{Python}. Crucially, we recorded the pure model generation latency, explicitly excluding time consumed by tool execution and network transmission. We then calculated the Pearson correlation between this generation time and both PTE and token counts. Sample data containing reasoning traces and timing statistics are released in our repository.

\begin{table*}[t]
    \centering
    \small
    \caption{Hardware Performance and Scaling Comparison on WebInstruct-Verified Trajectories}
    \label{tab:hardware_comparison}
    \begin{tabular}{lccccc}
        \toprule
        \textbf{Hardware} & \textbf{Peak TFLOPS} & \textbf{Memory BW} & \textbf{HOI} & \textbf{Scaling Factor} & \textbf{Spearman's Corr}\\
         & (FP16) & (TB/s) & (FLOPs/Byte) & ($\alpha$) & ($\rho$)\\
        \midrule
        NVIDIA H100  (Ref) & 1,513 & 2.00 & 756.5 & $1.0\times$ & 1.000000 \\
        NVIDIA H200        & 1,617 & 4.80 & 348.1 & $0.46\times$& 0.994505 \\
        NVIDIA A100        & 624   & 1.93 & 322.5 & $0.43\times$& 0.989011 \\
        NVIDIA V100        & 125   & 0.90 & 138.9 & $0.18\times$& 0.956044 \\
        NVIDIA RTX 4090    & 330   & 1.00 & 327.4 & $0.43\times$& 0.989011 \\
        \bottomrule
    \end{tabular}
\end{table*}

\subsection{Prompt Details}
An example system prompt that combines the three tools in the Webinstruct-Verified dataset is given in Fig. \ref{fig:sysprompt}. At the end of each rollout trajectory, DeepSeek-V3 served as a LLM judge to check answer correctness. Its prompt is shown in  Fig.\ref{fig:llmjudge}.

\section{Comprehensive Empirical Results}
\subsection{Full Benchmark Performance}
\label{sec:results}
This section presents the full experimental results for all evaluated models across the five benchmarks. We report the detailed metrics including Top-1 Accuracy, average uncached token counts (sum of prefill and decode), tool-usage times, and the overall Average PTE. These values correspond to the aggregated performance analysis discussed in Sec. \ref{sec:mainresults}. The complete numerical results are listed in Tab. \ref{tab1_part1} and Tab. \ref{tab1_part2}.
\begin{table*}[ht]
\begin{threeparttable}
\caption{Accuracy (\%), average uncached total tokens(prefil and decode), tooluse times, and average PTE comparison (Part 1: MATH \& AIME)}
\label{tab1_part1}
\footnotesize
\centering

\setlength{\tabcolsep}{2pt}

\begin{tabularx}{\textwidth}{>{\raggedright\arraybackslash}X *{12}{c}}
\toprule
\textbf{Models} & \multicolumn{4}{c}{\textbf{MATH500}} & \multicolumn{4}{c}{\textbf{AIME24}} & \multicolumn{4}{c}{\textbf{AIME25}} \\

\cmidrule(lr){2-5} \cmidrule(lr){6-9} \cmidrule(lr){10-13}

& acc$\uparrow$ & tokens$\downarrow$ & tooluse$\downarrow$ & PTE$\downarrow$ & acc$\uparrow$ & tokens$\downarrow$ & tooluse$\downarrow$ & PTE$\downarrow$ & acc$\uparrow$ & tokens$\downarrow$ & tooluse$\downarrow$ & PTE$\downarrow$ \\
\midrule

\rowcolor{AliceBlue}
\multicolumn{13}{l}{\textit{\textbf{Open-Source Dense Models}}} \\
Qwen2.5-7B-Instruct & 40.8 & 1808 & 1.178 & 2117 & 3.3 & 5086 & 0.500 & 5519 & 3.3 & 1461 & 0.133 & 1793 \\
Qwen2.5-32B-Instruct & 69.2 & 1517 & 1.840 & 1659 & 26.7 & 2040 & 1.267 & 2436 & 20.0 & 3382 & 2.467 & 3894 \\
Qwen2.5-72B-Instruct & 63.2 & 1668 & 1.264 & 1544 & 16.7 & 2855 & 1.267 & 2801 & 10.0 & 3237 & 1.367 & 3402 \\
Qwen3-32B & 55.6 & 7852 & 2.064 & 10912 & 53.3 & 27290 & 2.367 & 60374 & 33.3 & 34853 & 2.567 & 65356 \\
Llama-3.1-8B-Instruct & 28.0 & 844 & 0.922 & 963 & 10.0 & 1239 & 1.033 & 1307 & 10.0 & 3459 & 1.333 & 4203 \\
Llama-3.1-70B-Instruct & 38.6 & 718 & 1.162 & 702 & 3.3 & 735 & 1.100 & 719 & 6.7 & 815 & 1.000 & 797 \\
\midrule

\rowcolor{MistyRose}
\multicolumn{13}{l}{\textit{\textbf{Open-Source MoE Models}}} \\
Qwen3-235B-Instruct & 79.2 & 3266 & 1.024 & 2861 & 76.7 & 9624 & 1.033 & 9618 & 63.3 & 13494 & 1.100 & 19956 \\
Qwen3-235B-Thinking & 83.2 & 7023 & 1.022 & 8406 & 83.3 & 25261 & 1.000 & 34735 & 80.0 & 26811 & 0.933 & 35829 \\
GLM-4.5-Air & 72.9 & 7982 & 1.766 & 11569 & 66.7 & 22717 & 2.500 & 30359 & 53.3 & 17836 & 1.633 & 24489 \\
GLM-4.5 & 45.6 & 3621 & 0.960 & 4331 & 20.0 & 6478 & 0.833 & 8270 & 10.0 & 6182 & 0.967 & 9555 \\
DeepSeek-V3.1-Term & 81.4 & 4210 & 1.908 & 28203 & 56.7 & 31982 & 2.800 & 292724 & 60.0 & 21492 & 2.933 & 21639 \\
GPT-OSS-120B & 78.8 & 6363 & 2.604 & 18306 & 24016 & 5439 & 5.867 & 87049 & 60.0 & 25958 & 6.333 & 89627 \\
Tongyi-Deepresearch & 77.6 & 7925 & 1.242 & 27387 & 60.0 & 28315 & 1.133 & 143937 & 56.7 & 23912 & 0.900 & 104796 \\
\bottomrule
\end{tabularx}
\begin{tablenotes}
  \small 
  \item[1] GPT-OSS-120B uses default medium reasoning level.
  \item[2] Qwen3-32B uses under default generate config, which enables thinking mode.
\end{tablenotes}

\end{threeparttable}
\end{table*}

\begin{table*}[ht]
\begin{threeparttable}
\caption{Accuracy (\%), average uncached total tokens(prefil and decode), tooluse times, and average PTE comparison (Part 2: SimpleQA \& Webinstrucut)}
\label{tab1_part2}
\footnotesize
\centering

\setlength{\tabcolsep}{6pt}

\begin{tabularx}{\textwidth}{>{\raggedright\arraybackslash}X *{8}{c}}
\toprule
\textbf{Models} & \multicolumn{4}{c}{\textbf{SimpleQA}} & \multicolumn{4}{c}{\textbf{WebInstruct-Verified}} \\

\cmidrule(lr){2-5} \cmidrule(lr){6-9}

& acc$\uparrow$ & tokens$\downarrow$ & tooluse$\downarrow$ & PTE$\downarrow$ & acc$\uparrow$ & tokens$\downarrow$ & tooluse$\downarrow$ & PTE$\downarrow$ \\
\midrule

\rowcolor{AliceBlue}
\multicolumn{9}{l}{\textit{\textbf{Open-Source Dense Models}}} \\
Qwen2.5-7B-Instruct & 70.1 & 3533 & 2.409 & 3674 & 10.5 & 2589 & 1.485 & 3368\\
Qwen2.5-32B-Instruct & 50.2 & 5300 & 4.020 & 5361 & 47.7 & 2953 & 1.799 & 3813\\
Qwen2.5-72B-Instruct & 89.2 & 5909 & 2.380 & 6006 & 36.5 & 3407 & 1.775 & 3557 \\
Qwen3-32B & 65.6 & 2903 & 0.985 & 15573 & 38.5 & 3883 & 1.885 & 14635\\
Llama-3.1-8B-Instruct & 38.4 & 1024 & 1.140 & 1085 & 14.1 & 1373 & 1.291 & 1764\\
Llama-3.1-70B-Instruct & 51.2 & 3083 & 1.720 & 3120 & 5.5 & 1154 & 1.120 & 1187\\
\midrule

\rowcolor{MistyRose}
\multicolumn{9}{l}{\textit{\textbf{Open-Source MoE Models}}} \\
Qwen3-235B-A22B-Instruct & 85.1 & 3166 & 2.325 & 3184 & 47.0 & 3831 & 1.400 & 15772\\
Qwen3-235B-A22B-Thinking & 81.7 & 6325 & 1.787 & 9306 & 45.0 & 9787 & 1.180 & 63154\\
GLM-4.5-Air & 84.6 & 17992 & 4.860 & 22045 & 19.0 & 4854 & 1.065 & 5170\\
GLM-4.5 & 92.9 & 17122 & 4.455 & 20617 & 13.5 & 2964 & 1.005 & 3424\\
DeepSeek-V3.1-Terminus & 87.6 & 20783 & 5.235 & 21023 & 43.6 & 26583 & 6.840 & 27137\\
GPT-OSS-120B & 90.4 & 12641 & 3.480 & 15474 & 45.0 & 10640 & 3.740 & 21256\\
Tongyi-Deepresearch & 4.8 & 14155 & 6.940 & 45677 & 21.5 & 16509 & 3.830 & 63154\\
\bottomrule
\end{tabularx}
\begin{tablenotes}
  \small 
  \item[1] GPT-OSS-120B uses default medium reasoning level.
  \item[2] Qwen3-32B uses under default generate config, which enables thinking mode.
\end{tablenotes}

\end{threeparttable}
\end{table*}

\subsection{PTE Analysis by Reasoning Correctness}
\label{sec:efficiency_results}
\begin{figure*}[t]
  \centering
  \small                         
  \begin{subfigure}[b]{0.48\linewidth}
    \centering
    \includegraphics[width=\linewidth]{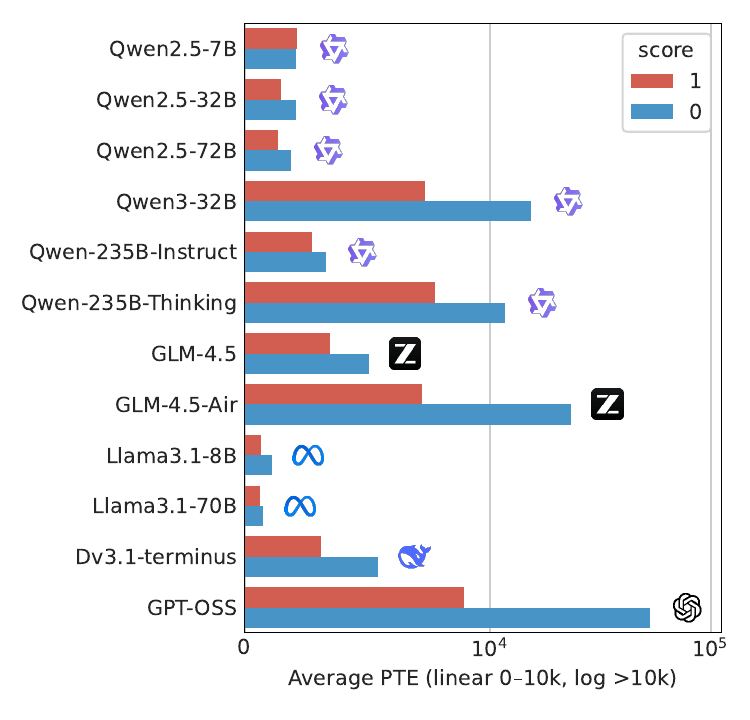}
    \caption{MATH500}
    \label{fig:bench1}
  \end{subfigure}
  \hfill                          
  \begin{subfigure}[b]{0.48\linewidth}
    \centering
    \includegraphics[width=\linewidth]{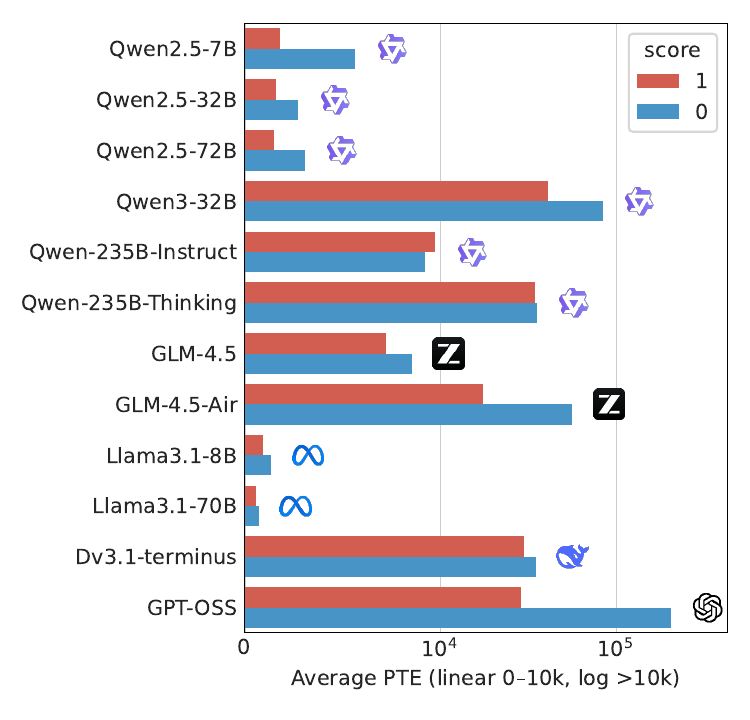}
    \caption{AIME24}
    \label{fig:bench2}
  \end{subfigure}

  \begin{subfigure}[b]{0.31\linewidth}
    \centering
    \includegraphics[width=\linewidth]{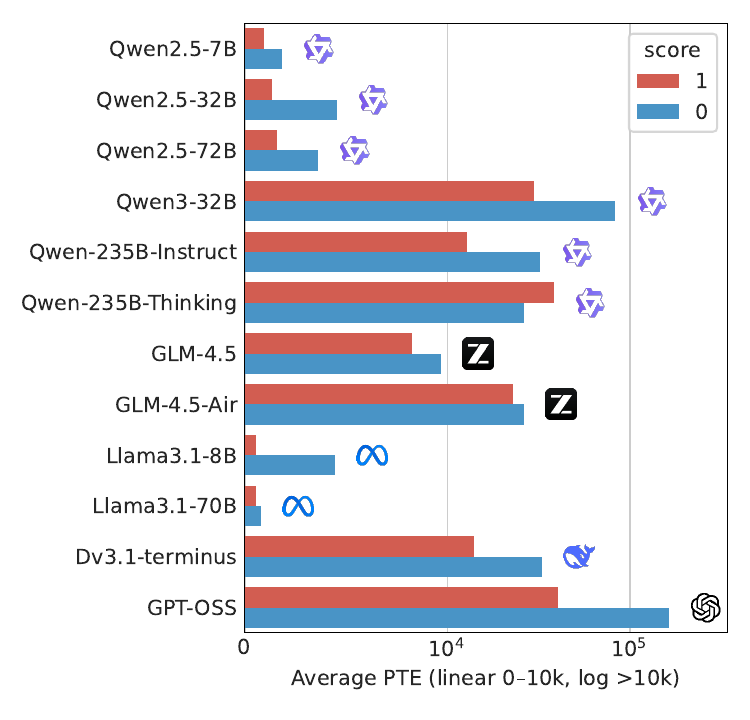}
    \caption{AIME25}
    \label{fig:bench3}
  \end{subfigure}
  \hfill
  \begin{subfigure}[b]{0.31\linewidth}
    \centering
    \includegraphics[width=\linewidth]{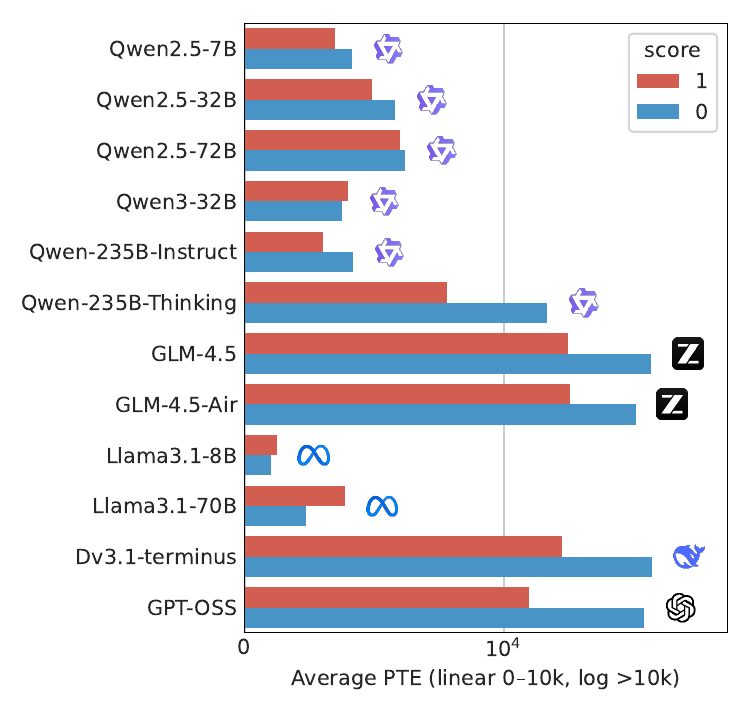}
    \caption{SimpleQA}
    \label{fig:bench4}
  \end{subfigure}
  \hfill
  \begin{subfigure}[b]{0.31\linewidth}
    \centering
    \includegraphics[width=\linewidth]{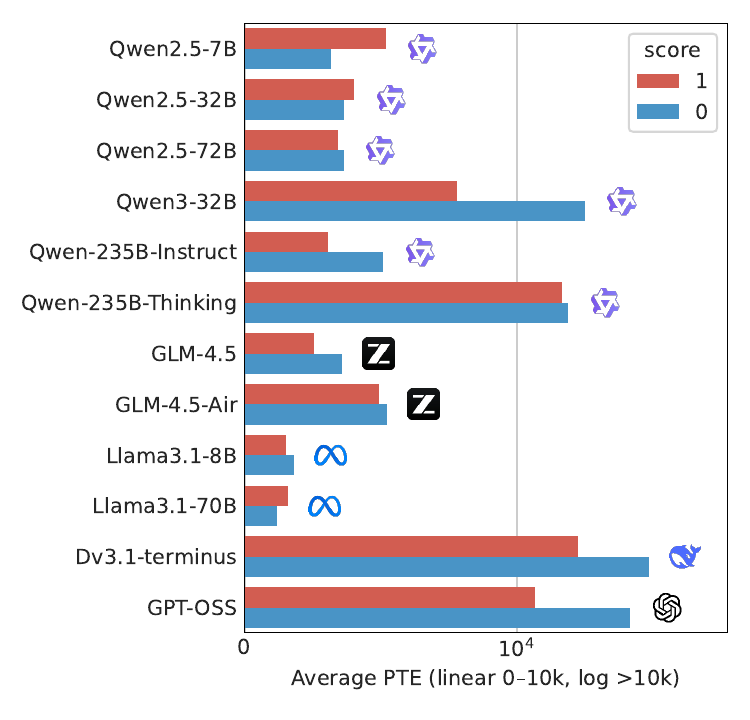}
    \caption{Webinstruct}
    \label{fig:bench5}
  \end{subfigure}

  \caption{Comparison of \textbf{PTE} between Correct(red) and Incorrect(blue) Reasoning Trajectories. Incorrect reasoning is consistently associated with higher PTE across models and tasks.}
  \label{fig:Correct vs Incorrect}
\end{figure*}

This section supplements the efficiency-accuracy analysis presented in Sec. \ref{sec:effi-acc}. While Fig. \ref{fig:score01vsPTE_simplified} provides an aggregated visualization of the cost distribution, Fig. \ref{fig:Correct vs Incorrect} presents the granular performance metrics for individual models.

\section{Qualitative Case Studies}
\subsection{Pattern I: Confirmatory Tool Usage}
\label{sec:appendix_Z}
This example illustrates the Confirmatory Tool Usage behavior discussed in Sec. \ref{sec:inefficient}, where Qwen3-235B-Thinking use python tool to verify its internal thoughts rather than using it as direct solvers. See Fig. \ref{fig: confirm traj}.

\vspace{-1em}
\begin{figure*}[!ht] 
\begin{AIbox}{Trajectory Example of Qwen3-235B-Thinking on AIME24}
\textbf{Question}: Jen enters a lottery by picking $4$ distinct numbers from $S=\{1,2,3,\cdots,9,10\}.$ $4$ numbers are randomly chosen from $S.$ She wins a prize if at least two of her numbers were $2$ of the randomly chosen numbers, and wins the grand prize if all four of her numbers were the randomly chosen numbers. The probability of her winning the grand prize given that she won a prize is $\tfrac{m}{n}$ where $m$ and $n$ are relatively prime positive integers. Find $m+n$.

\textbf{Groundtruth}(invisible to model): 116

\tcblower
{\color{deepblue}\bf \large Assistant:} (use \textbf{python} tool to verify its internal thoughts rather than using it as direct solvers) 
\begin{lstlisting}[style=prompt]
Okay, let's try to solve this problem step by step. First, let's make sure I understand the problem correctly.\n\nJen picks 4 distinct numbers from the set S = {1, 2,..., 10}. Then, 4 numbers are randomly chosen from S (let's call this the winning set). She wins a prize if at least two of her numbers match the winning set. She wins the grand prize if all four match. We need to find the probability of her winning the grand prize given that she won a prize. That is, P(grand prize | won a prize) = P(grand prize) / P(won a prize), since grand prize is a subset of winning a prize.\n\nFirst, let's recall that in probability, the conditional probability P(A|B) = P(A \cap B) / P(B).

......

Exactly 0 matches: C(2,0)*C(3,2)=1*3=3.\n\nPrize-winning sets (at least 1 match): 6+1=7.\n\nConditional probability: 1/7. Which makes sense. If we compute it as (1/10)/(7/10)=1/7. Correct.\n\nSo the method is sound.\n\nBack to the original problem: 115 is the number of prize-winning sets, 1 is grand prize. So 1/115.\n\nWait, but let's compute C(10,4) to be sure: 10*9*8*7/(4*3*2*1)=210. Correct.\n\nC(4,2)=6, C(6,2)=15, 6*15=90. Correct.\n\nC(4,3)=4, C(6,1)=6, 4*6=24. Correct.\n\nC(4,4)=1, C(6,0)=1. Correct.\n\n90+24+1=115. Correct.\n\nSo the conditional probability is 1/115. Therefore, m=1, n=115, m+n=116.\n\nI think that's it. Let me just write a Python code to verify the counts.\n\nWe can compute:\n\ntotal = comb(10, 4)\n\ngrand = comb(4, 4) * comb(6, 0)\n\nprize = comb(4, 2)*comb(6, 2) + comb(4, 3)*comb(6, 1) + comb(4, 4)*comb(6, 0)\n\nprobability = grand / prize\n\nBut since we need the fraction, it's grand / prize = 1 / 115.\n\nLet's code it using math.comb (since Python 3.10+ has math.comb, but if not, we can define it).\n\nWait, in Python, math.comb is available from 3.8 onwards. Let's assume the environment supports it.\n\nCode:\n\nimport math\n\n# Total number of winning sets (not needed for the ratio, but for verification)\ntotal = math.comb(10, 4)  # 210\n\n# Number of grand prize sets\ngrand = math.comb(4, 4) * math.comb(6, 0)  # 1*1=1\n\n# Number of prize-winning sets (at least 2 matches)\nprize_2 = math.comb(4, 2) * math.comb(6, 2)  # 6*15=90\nprize_3 = math.comb(4, 3) * math.comb(6, 1)  # 4*6=24\nprize_4 = grand  # 1\nprize_total = prize_2 + prize_3 + prize_4  # 90+24+1=115\n\n# Conditional probability\nprob = grand / prize_total\n\n# But we need the fraction m/n where m and n are coprime\nfrom fractions import Fraction\nfraction = Fraction(grand, prize_total)\nm = fraction.numerator\nn = fraction.denominator\nprint(f\"m: {m}, n: {n}, m+n: {m+n}\")\n\nThis should output m=1, n=115, m+n=116.\n\nLet me check if math.comb(6,2) is 15: 6*5/2=15, yes. math.comb(4,2)=6, correct.\n\nYes, this code should confirm the counts.\n

"tool_calls": {"arguments": "{\"code\": \"import math\\nfrom fractions import Fraction\\n\\ntotal = math.comb(10, 4)\\ngrand = math.comb(4, 4) * math.comb(6, 0)\\nprize_2 = math.comb(4, 2) * math.comb(6, 2)\\nprize_3 = math.comb(4, 3) * math.comb(6, 1)\\nprize_4 = grand\\nprize_total = prize_2 + prize_3 + prize_4\\nfraction = Fraction(grand, prize_total)\\nm = fraction.numerator\\nn = fraction.denominator\\nprint(f\\\"m: {m}, n: {n}, m+n: {m+n}\\\")\"}", "name": "python_tool"}, "type": "function"}]
\end{lstlisting}

{\color{RosyBrown}\bf Tool:} (tool response)
\begin{lstlisting}[style=prompt]
"stdout:\nm: 1, n: 115, m+n: 116\n\nstderr:"
\end{lstlisting}

\end{AIbox} 
\caption{\textbf{Pattern I: Confirmatory Tool Usage.} An example from AIME24 where Qwen3-235B-Thinking solves the problem internally first, arriving at the correct answer. However, it subsequently invokes the Python tool solely to verify this known result. This "Think-then-Verify" behavior unnecessarily inflates the context length and PTE cost without contributing new information to the solution.}
\label{fig: confirm traj}
\vspace{-1em}
\end{figure*}

\subsection{Pattern II: Tool-Mixing Example Trajectory}
\label{sec:appendix_A}
This example illustrates the "tool-mixing" behavior discussed in Sec. \ref{sec:inefficient}, where DeepSeek-V3.1-Terminus alternates between search and Python tools within a single trajectory on the Webinstruct-Verified benchmark. See Fig. \ref{fig: mixing traj}.

\vspace{-1em}
\begin{figure*}[!ht] 
\begin{adjustbox}{width=0.95\textwidth} 
    \begin{minipage}{\textwidth} 
\begin{AIbox}{Trajectory Example of DeepSeek-V3.1-Terminus on Webinstruct-Verified}
\textbf{Question}: Organic agriculture implies farming techniques that use no \underline{\hspace{1cm}}.

\textbf{Groundtruth}(invisible to model): Insecticides. Inorganic or synthetic fertilizers. Herbicides.
\tcblower
{\color{deepblue}\bf \large Assistant:} (use \textbf{search} tool) 
\begin{lstlisting}[style=prompt]
I'll help you understand what organic agriculture excludes by searching for information about organic farming techniques.

"tool_calls": {"arguments": "{\"query\": [\"organic agriculture definition what is not used\", \"organic farming prohibited substances techniques\", \"what does organic agriculture exclude\"]}", "name": "google_search_tool"}

\end{lstlisting}

{\color{RosyBrown}\bf Tool:} (tool response)
\begin{lstlisting}[style=prompt]
"A Google search for 'organic agriculture definition what is not used' found 10 results:\n\n## Web Results

1. [5 Reasons Why Organic Farming Is Not Sustainable]
(https://www.edengreen.com/blog-collection/organic-farming-sustainability)
Organic produce is grown without synthetic pesticides, fertilizers, or GMOs. It must meet strict USDA certification standards. These standards include using ...See more

2. [What Organic Farming (and Processing) Doesn't Allow]
(https://www.usda.gov/about-usda/news/blog/organic-101-what-organic-farming-and-processing...)
Methods like irradiation, sewage sludge, and genetic engineering are all expressly prohibited from being used when growing or processing organic foods.See more

3. [Organic vs. Conventional Farming]..."
\end{lstlisting}

{\color{deepblue}\bf \large Assistant:} (use \textbf{visit} tool) 
\begin{lstlisting}[style=prompt]
Now let me visit the most relevant USDA page to get detailed information about what organic agriculture excludes:

"tool_calls": {"arguments": "{\"url\": \"https://www.usda.gov/about-usda/news/blog/organic-101-what-organic-farming-and-processing...\", \"goal\": \"Find specific information about what substances and methods are prohibited in organic agriculture\"}", "name": "visit_tool"}
\end{lstlisting}

{\color{RosyBrown}\bf Tool:} (tool response)
\begin{lstlisting}[style=prompt]
The useful information in https://www.usda.gov/about-usda/news/blog/organic-101-what-organic-farming-and-processing... for user goal Find specific information about what substances and methods are prohibited in organic agriculture as follows: \n\nEvidence in page: \nA very important part of the process-based regulatory framework is the prohibition of certain methods in organic production and handling...
\end{lstlisting}

{\color{deepblue}\bf \large Assistant:}
\begin{lstlisting}[style=prompt]
......
\end{lstlisting}

{\color{RosyBrown}\bf \large Tool:}
\begin{lstlisting}[style=prompt]
......
\end{lstlisting}

{\color{deepblue}\bf \large Assistant:}(use \textbf{python} tool) 
\begin{lstlisting}[style=prompt]
Based on my research, I can now provide a comprehensive answer about what organic agriculture excludes. Let me use Python to verify and organize the key points:

"tool_calls": {"arguments": "{\"code\": \"# Organizing the key prohibited substances and methods in organic agriculture\\n\\nprohibited_items = [\\n    \\\"synthetic fertilizers\\\",\\n    \\\"synthetic pesticides\\\",\\n    \\\"genetically modified organisms (GMOs)\\\",\\n    \\\"irradiation\\\",\\n    \\\"sewage sludge\\\",\\n    \\\"artificial colors\\\",\\n    \\\"artificial preservatives\\\", \\n    \\\"artificial flavors\\\",\\n    \\\"genetic engineering methods (cell fusion, recombinant DNA, etc.)\\\"\\n]\\n\\nprint(\\\"Organic agriculture excludes the use of:\\\")\\nfor i, item in enumerate(prohibited_items, 1):\\n    print(f\\\"{i}. {item}\\\")\\n\\nprint(\\\"\\\\nThe basic principle is: 'allow natural substances and prohibit synthetic substances'\\\")\\nprint(\\\"\\\\nSource: USDA National Organic Program standards and EPA guidelines\\\")\"}", "name": "python_tool"}
\end{lstlisting}

{\color{RosyBrown}\bf \large Tool:} (tool response)
\begin{lstlisting}[style=prompt]
"stdout:\nOrganic agriculture excludes the use of:\n1. synthetic fertilizers\n2. synthetic pesticides\n3. genetically modified organisms (GMOs)\n4. irradiation\n5. sewage sludge\n6. artificial colors\n7. artificial preservatives\n8. artificial flavors\n9. genetic engineering methods (cell fusion, recombinant DNA, etc.)\n\nThe basic principle is: 'allow natural substances and prohibit synthetic substances'\n\nSource: USDA National Organic Program standards and EPA guidelines\n\nstderr:"
\end{lstlisting}

{\color{deepblue}\bf \large Assistant:}(final answer \textcolor{ForestGreen}{\ding{51}}) 
\begin{lstlisting}[style=prompt]
<ANSWER>Organic agriculture implies farming techniques that use no synthetic fertilizers, synthetic pesticides, genetically modified organisms (GMOs), irradiation, sewage sludge, artificial colors, artificial preservatives, artificial flavors, or genetic engineering methods. The fundamental principle is to allow natural substances while prohibiting synthetic ones.</ANSWER>
\end{lstlisting}

\end{AIbox} 
\end{minipage}
\end{adjustbox}
\caption{\textbf{Pattern II: Tool-Mixing.} An example from WebInstruct-Verified where DeepSeek-V3.1-Terminus fragmentally alternates between Search and Python tools. This behavior accumulates context of intermediate outputs and inflates the PTE cost, yet yields no obvious accuracy improvement compared to single-toolset strategies (as evidenced in Fig. \ref{fig:bench5_1}).}
\label{fig: mixing traj}
\vspace{-1em}
\end{figure*}

\subsection{Pattern III: Lack of Tool Priors}
\label{sec:appendix_B}
This example supports the analysis in Sec. \ref{sec:inefficient}. It shows a trajectory from Qwen-2.5-7B-Instruct on AIME24 benchmark, where the provision of a \texttt{Python} tool (for which it lacks strong pre-training) leads to an inefficient and incorrect reasoning path. This contrasts with specialized code-reasoning methods \cite{chen2025codedrivennumbersequencecalculation} that optimize the inductive reasoning abilities of LLMs through dedicated code-driven training, suggesting that a tool is most efficient when the model has already learned how to use it. See Fig. \ref{fig: prior traj}.

\vspace{-1em}
\begin{figure*}[!ht] 
\begin{AIbox}{Trajectory Example of Qwen-2.5-7B-Instruct on AIME24}
\textbf{Question}: Find the largest possible real part of $[(75+117i)z+\frac{96+144i}{z}]$ where $z$ is a complex number with $|z|=4$.

\textbf{Groundtruth}(invisible to model): 540
\tcblower
{\color{deepblue}\bf \large Assistant:} (\textbf{python} tool invoked, but no print function used) 
\begin{lstlisting}[style=prompt]
To find the largest possible real part of the expression \\((75+117i)z + \\frac{96+144i}{z}\\) where \\(z\\) is a complex number with \\(|z|=4\\), we can start by expressing \\(z\\) in its polar form. Since \\(|z|=4\\), we can write \\(z = 4e^{i\\theta}\\) for some angle \\(\\theta\\).\n\nLet's denote \\(z = 4(\\cos\\theta + i\\sin\\theta)\\). Then we have:\n\\[\n(75+117i)z = (75+117i)4(\\cos\\theta + i\\sin\\theta) = 4(75\\cos\\theta - 117\\sin\\theta) + i(75\\sin\\theta + 117\\cos\\theta)\n\\]\nand\n\\[\n\\frac{96+144i}{z} = \\frac{96+144i}{4(\\cos\\theta + i\\sin\\theta)} = \\frac{1}{4}(96\\cos\\theta - 144\\sin\\theta) + i(96\\sin\\theta + 144\\cos\\theta).\n\\]\nAdding these two expressions together, we get:\n\\[\n(75+117i)z + \\frac{96+144i}{z} = 4(75\\cos\\theta - 117\\sin\\theta) + \\frac{1}{4}(96\\cos\\theta - 144\\sin\\theta) + i\\left(75\\sin\\theta + 117\\cos\\theta + 96\\sin\\theta + 144\\cos\\theta\\right).\n\\]\nThe real part of this expression is:\n\\[\n4(75\\cos\\theta - 117\\sin\\theta) + \\frac{1}{4}(96\\cos\\theta - 144\\sin\\theta) = 300\\cos\\theta - 468\\sin\\theta + 24\\cos\\theta - 36\\sin\\theta = 324\\cos\\theta - 504\\sin\\theta.\n\\]\nWe need to maximize the expression \\(324\\cos\\theta - 504\\sin\\theta\\). This is a linear combination of sine and cosine, and its maximum value is the amplitude of the function, which is \\(\\sqrt{324^2 + 504^2}\\).\n\nLet's calculate this amplitude:\n\\[\n\\sqrt{324^2 + 504^2} = \\sqrt{104976 + 254016} = \\sqrt{358992} = 36\\sqrt{279}.\n\\]\nSo the maximum value of the real part is \\(36\\sqrt{279}\\). We can simplify \\(\\sqrt{279}\\) as \\(\\sqrt{9 \\cdot 31} = 3\\sqrt{31}\\), so the maximum value is:\n\\[\n36 \\cdot 3\\sqrt{31} = 108\\sqrt{31}.\n\\]\n\nLet's verify this using Python code.

"tool_calls": {"arguments": "{\"code\": \"import math\\n# Calculate the maximum value\\nmax_value = 36 * 3 * math.sqrt(31)\\nmax_value\"}", "name": "python_tool"}
\end{lstlisting}

{\color{RosyBrown}\bf Tool:} (tool response)
\begin{lstlisting}[style=prompt]
"stdout:\n\nstderr:"
\end{lstlisting}

{\color{deepblue}\bf \large Assistant:}(final answer\textcolor{red}{\ding{55}}) 
\begin{lstlisting}[style=prompt]
The maximum value of the real part of the expression \\((75+117i)z + \\frac{96+144i}{z}\\) where \\(z\\) is a complex number with \\(|z|=4\\) is \\(108\\sqrt{31}\\).\n\nThus, the final answer is:\n\\[\n\\boxed{108\\sqrt{31}}\n\\]
\end{lstlisting}

\end{AIbox} 
\caption{\textbf{Pattern III: Lack of Tool Priors.} An example from AIME24 where Qwen-2.5-7B-Instruct fails to utilize the Python tool effectively. The model invokes the code interpreter but forgets to include a print statement, resulting in an empty output. This suggests a lack of pretraining exposure to the tool environment, leading to wasted inference steps.}
\label{fig: prior traj}
\vspace{-1em}
\end{figure*}

\subsection{Pattern IV: Tool Format Collapse}
\label{sec:appendix_C}
This example illustrates the sensitivity to tool format phenomenon discussed in Sec. \ref{sec:inefficient}. It shows a SimpleQA trajectory where Tongyi-Deepresearch fails to correctly invoke a tool when presented with a semantically identical but syntactically different tool definition from what it was trained on. See Fig. \ref{fig: format traj}.

\vspace{-1em}
\begin{figure*}[!ht] 
\begin{AIbox}{Trajectory Example of Tongyi-Deepresearch on SimpleQA}
\textbf{Question}: What is the British-American kickboxer Andrew Tate's kickboxing name?

\textbf{Groundtruth}(invisible to model): King cobra

{\color{OliveGreen}\bf \large Tool Config:}(Notice that we define \textbf{google\_search\_tool} and \textbf{visit\_tool})
\begin{lstlisting}[style=prompt]
"tools":[{"type": "function", "function": {"name": "google_search_tool", "description": "Performs batched web searches: supply a list of queries; the tool retrieves the top 10 results for each query in one call, each including a title, URL, and abstract. Use these abstracts to evaluate the relevance of each result and determine which URL to visit for more detailed information based on the original query.", "parameters": {"type": "object", "properties": {"query": {"type": "array", "items": {"type": "string"}, "description": "Array of query strings. Include multiple complementary search queries in a single call."}}, "required": ["query"]}}}, 

{"type": "function", "function": {"name": "visit_tool", "description": "Visit a single webpage URL and return a concise summary of the content tailored to the goal provided. The goal should be directly derived from the original question to ensure the summary extracts only the most relevant information. The summary should include key facts, dates, names, or events as applicable.", "parameters": {"type": "object", "properties": {"url": {"type": "string", "description": "The URL of the webpage to visit."}, "goal": {"type": "string", "description": "The goal of the visit for the webpage."}}, "required": ["url", "goal"]}}}]
\end{lstlisting}

\tcblower
{\color{deepblue}\bf \large Assistant:} (use \textbf{search} tool, which is not defined in our framework but provided in the Tongyi opensource framework) 
\begin{lstlisting}[style=prompt]
The user asks: \"What is the British-American kickboxer Andrew Tate's kickboxing name?\" They refer to Andrew Tate as a British-American kickboxer. His kickboxing name is \"User83\"? Or something? Let's recall. Andrew Tate, also known as \"Croatian\" or \"Andrew Tate\" is known as \"Cairo\"?? I'm not sure. Let's recall specifics: Andrew Tate, born 1 December 1986, is a British-American internet personality, businessman, and former professional kickboxer. He has a nickname in kickboxing: \"Croatian Wolf\"? Actually a known nickname is \"Crock\"? Let's confirm.\n\nI recall various references: Andrew Tate's actual nickname in kickboxing is \"Crock\"? I might be mixing with other things.\n\nLet's recall known facts: Andrew Tate's kickboxing nickname is \"Crockett\"? I've seen his social media profile uses the handle \"Crock\"? Let's search."

"tool_calls": {"arguments": "{\"query\": [\"Andrew Tate kickboxing name nickname\", \"Andrew Tate kickboxing moniker\", \"What is Andrew Tate's kickboxing name?\", \"Andrew Tate fight name\", \"Andrew Tate kickboxing record nickname\"]}", "name": "search"}, "type": "function"}]
\end{lstlisting}

{\color{RosyBrown}\bf Tool:} (tool response)
\begin{lstlisting}[style=prompt]
"Error: tool 'search' not registerd."
\end{lstlisting}

{\color{deepblue}\bf \large Assistant:} (use \textbf{search} tool again, fail to be parsed by engine due to extra linefeed, final answer\textcolor{red}{\ding{55}}) 
\begin{lstlisting}[style=prompt]
"<think>\nThe tool name is \"search\" under the \"functions\" namespace. Usage: we need to provide a JSON with key \"name\": \"search\", \"arguments\": ...? The conversation format for calling the tool is as per the instructions: we need to output:\n\n<tool_call>\n{\"name\": \"search\", \"arguments\": {\"query\": [...]}}\n</tool_call>\n\nMake sure the name matches the function defined: search. So do that.\n</think>\n\n<tool_call>\n{\"name\": \"search\", \"arguments\": {\"query\": [\"Andrew Tate kickboxing name nickname\", \"Andrew Tate kickboxing moniker\", \"What is Andrew Tate's kickboxing name?\", \"Andrew Tate fight name\", \"Andrew Tate kickboxing record nickname\"]}}\n</tool_call>"
\end{lstlisting}

\end{AIbox} 
\caption{\textbf{Pattern IV: Tool Format Collapse.} An example from SimpleQA illustrating model brittleness. Tongyi-Deepresearch fails to adhere to the predefined tool schema, hallucinating a search tool format that differs syntactically from the system prompt (likely reverting to its training data format). This mismatch causes immediate parsing error and recurring failures in the following steps despite the semantic intent being correct.}
\label{fig: format traj}
\vspace{-1em}
\end{figure*}

\section{Quantitative Analysis of Inefficiency Patterns}
\label{app:quantitative-patterns}

We measured the exact PTE cost of each inefficiency pattern. Although these patterns can co-occur, they appear in distinct model-task settings (Sec. \ref{sec:inefficient}) and do not always occur together. We therefore quantified each within its primary occurrence setting (model-task pair), where per-pattern overhead remains clearly measurable.

We defined heuristic rules to detect each pattern and measured the \textbf{Frequency} (\% of trajectories exhibiting the pattern) and \textbf{Cost Multiplier} (PTE increase compared to pattern-free trajectories in the same setting). As shown in Tab. \ref{tab:pattern-costs}, these patterns are not merely qualitative quirks but significant drivers of computational inefficiency, with cost multipliers ranging from $1.77\times$ to $2.42\times$.

\begin{table*}[ht]
\caption{Quantitative breakdown of inefficiency patterns. Frequency and cost multiplier measured within dominant model-task settings where each pattern occurs most distinctly.}
\centering
\small
\begin{tabular}{@{}p{2.3cm}p{3.8cm}p{3.4cm}cc@{}}
\toprule
\textbf{Pattern} & \textbf{Dominant Setting} & \textbf{Heuristic Definition} & \textbf{Freq.} & \textbf{Cost} \\
\midrule
Confirmatory & Qwen3-235B-Instruct (MATH500) & Answer token generated pre-tool & 81\% & 1.77$\times$ \\
Tool-Mixing & DeepSeek-V3.1-Term (WebInstruct) & Unique tool types $>1$ & 59\% & 2.42$\times$ \\
Format Collapse & Tongyi-DeepResearch (SimpleQA) & Schema/JSON Error in output & 100\% & N/A \\
Lack of Priors & Qwen2.5-72B-Instruct (AIME24) & Empty or Error tool return & 33\% & 2.15$\times$ \\
\bottomrule
\end{tabular}
\label{tab:pattern-costs}
\end{table*}

\paragraph{Detection Heuristics.} 
\textbf{Confirmatory:} Detected when the model generates an answer or final value in its reasoning text before invoking the Python tool for verification. \textbf{Tool-Mixing:} Detected when a trajectory uses more than one distinct tool type (e.g., both search and Python). \textbf{Format Collapse:} Detected when the output contains JSON parsing errors or schema violations. \textbf{Lack of Priors:} Detected when a tool call returns empty output or execution error due to missing print statements or incorrect syntax.

\section{PTE reflects reasoning inefficiency beyond query difficulty}
\label{app:pte-inefficiency}

To verify that high PTE reflects reasoning inefficiency rather than inherent problem difficulty, we conduct three stratified analyses on MATH500 (dataset-provided levels 1–5) and AIME25 (GPT-4o-annotated levels 4–5).

\subsection{Intra-Level Analysis: Efficiency gaps persist when difficulty is controlled}
If PTE primarily reflected question difficulty, success and failure at the same difficulty level would be expected to yield similar PTE distributions. Contradicting this hypothesis, incorrect trajectories exhibit significantly higher PTE than correct ones within every difficulty level (See Tab. \ref{tab:intra-level}).

\begin{table*}[ht]
\caption{Percentage PTE increase of incorrect trajectories over correct ones within the same difficulty level.}
\centering
\small
\begin{tabular}{lccc}
\toprule
\textbf{Model} & \textbf{Level 1 Gap} & \textbf{Level 3 Gap} & \textbf{Level 5 Gap} \\
\midrule
GPT-OSS & +5.1\% & +158.2\% & +371.1\% \\
GLM-4.5-Air & +419.9\% & +82.7\% & +259.9\% \\
Qwen3-32B & +793.2\% & +69.6\% & +100.2\% \\
Qwen2.5-72B & +43.0\% & +32.3\% & +16.9\% \\
Dv3.1-terminus & +341.8\% & -0.2\% & +46.3\% \\
\bottomrule
\end{tabular}
\label{tab:intra-level}
\end{table*}

By holding difficulty constant, these gaps suggest that reasoning efficiency remains a major source of PTE variation beyond difficulty. Notably, the largest occur at Level 1 (up to 793\%), where failures are often associated with reasoning inefficiencies, as detailed in Section \ref{sec:inefficient} and Appendix \ref{sec:appendix_Z}.

\subsection{Inter-Level Analysis: Efficiency dominates difficulty in determining cost}
If PTE reflected only difficulty, same-level success and failure would show similar PTEs. Instead, failures consistently exhibit higher PTE across all levels (See Tab. \ref{tab:inter-level}).

\begin{table*}[ht]
\caption{Percentage of Level-5 correct trajectories with lower PTE than average failures at easier levels.}
\centering
\small
\begin{tabular}{lcccc}
\toprule
\textbf{Dataset} & \textbf{L5 Correct vs L1 Wrong} & \textbf{L5 Correct vs L2 Wrong} & \textbf{L5 Correct vs L3 Wrong} & \textbf{L5 Correct vs L4 Wrong} \\
\midrule
MATH500 & 57.6\% (n=109) & 38.4\% (n=290) & 57.7\% (n=475) & 68.4\% (n=605) \\
AIME25 & N/A & N/A & N/A & 34.4\% (n=126) \\
\bottomrule
\end{tabular}
\label{tab:inter-level}
\begin{tablenotes}
  \small 
  \item[1] AIME25 problems were only annotated as Level 4--5 by GPT-4o, despite the 1--5 scale.
\end{tablenotes}
\end{table*}

Beyond aggregate trends, this holds within individual models: GPT-OSS uses 75\% less PTE to correctly solve a Level-5 problem (24,302) than to fail on a Level-4 problem (97,183). This shows that reasoning efficiency can frequently outweigh intrinsic problem difficulty in determining overall PTE.

\subsection{Statistical Isolation: PTE predicts accuracy independently of difficulty}
Partial correlation (controlling for difficulty) reveals a significant negative PTE-accuracy association ($r = -0.040, p = 0.002$), indicating the relationship persists beyond difficulty effects. In conclusion, PTE reflects reasoning inefficiency beyond query difficulty.

\section{Comparison with Commercial API Pricing}
\label{app:api-pricing}

We compare PTE against commercial API pricing metrics to demonstrate its superiority in capturing non-linear hardware bottlenecks. We calculated API Cost using the standard formula:
\[
\text{Cost} = {Token}_{\text{in}} \times P_{\text{in}} + {Token}_{\text{out}} \times P_{\text{out}},
\]
where input tokens include the full context history (assuming cache miss) to reflect real-world context accumulation costs. We selected three representative industry price ratios ($P_{\text{in}}:P_{\text{out}}$): 1:1.5 (DeepSeek-V3.2), 1:3 (Standard), and 1:4 (GPT-4o/Claude 3.5).

As shown in Table~\ref{tab:pricing-comparison}, PTE achieves the highest correlation with wall-clock latency ($r=0.925$) because it models actual hardware bottlenecks (memory bandwidth constraints), not arbitrary commercial pricing strategies. Its hardware-adaptive design ensures robustness across any deployment scenario.

\begin{table*}[ht]
\caption{Correlation with wall-clock latency: PTE vs. token-count and commercial pricing baselines.}
\centering
\small
\begin{tabular}{@{}lccl@{}}
\toprule
\textbf{Metric} & \textbf{Ratio (In:Out)} & \textbf{Pearson $r$} & \textbf{Notes} \\
\midrule
Naive Token Count & 1:1 & 0.625 & Ignores context accumulation \\
Pricing (DeepSeek-V3.2) & 1:1.5 & 0.681 & Fixed I/O price ratio \\
Pricing (Standard) & 1:3 & 0.734 & Fixed I/O price ratio \\
Pricing (GPT-4o/Claude-3.5) & 1:4 & 0.758 & Fixed I/O price ratio \\
\midrule
\textbf{PTE (Ours)} & Hardware-Adaptive & \textbf{0.925} & Captures physical bandwidth bottlenecks \\
\bottomrule
\end{tabular}
\label{tab:pricing-comparison}
\end{table*}

\end{document}